\documentclass[aps,prb,amsmath,twocolumn,amssymb,floatfixng,showpacs,
superscriptaddress,footinbib]{revtex4-1}
\pdfoutput=1
\usepackage[dvips]{graphics}
\usepackage{bm}
\usepackage{epsfig}
\usepackage{enumerate}
\usepackage{subfigure}
\usepackage{color}
\usepackage{graphicx}
\usepackage{hyperref}
\hypersetup{
    pdfnewwindow=true,      
    colorlinks=true,       
    linkcolor=magenta,          
    citecolor=blue,        
    filecolor=blue,      
    urlcolor=blue        
}

\newcommand{\etal}{{\it et al.~}}
\newcommand{\ie}{{\it i.e., }}

\begin{document} 
\title{Valley-polarized magnetoconductivity and particle-hole symmetry breaking in a periodically modulated $\alpha$-$\mathcal{T}_3$ lattice} 

\author{SK Firoz Islam}
\email{firoz@iopb.res.in}
\author{Paramita Dutta}
\email{paramitad@iopb.res.in}
\thanks{Both the authors have contributed equally to this work.}
\affiliation{Institute of Physics, Sachivalaya Marg, Bhubaneswar-751005, India}

\begin{abstract}
We explore the transport properties of a periodically modulated $\alpha$-$\mathcal{T}_3$ lattice in the presence of a perpendicular magnetic 
field. The effect of the Berry phase on electrical conductivity oscillation, so-called Weiss oscillation, caused by the modulation 
induced non-zero drift velocity of charge carriers is investigated. Employing linear response theory within the low temperature 
regime, we analyze Weiss oscillation as a function of the external magnetic field for both electrically and magnetically modulated 
$\rm \alpha$-$\mathcal{T}_3$ lattice numerically as well as analytically. The Berry phase makes this hexagonal lattice structure behave
differently than other two-dimensional fermionic systems. It causes a significant valley polarization in magnetoconductivity. Most 
interestingly, the combined effect of both modulations breaks the particle-hole symmetry and causes a smooth transition from even 
(odd) to odd (even) filling fraction corresponding to the density of states peaks by means of the Berry phase. 
\end{abstract}

\maketitle
\section{Introduction}
Since the discovery of the most celebrated atomically thin material, graphene~\cite{novo,neto},
the search for new Dirac materials has been growing day by day owing to their peculiar electronic structure
and possible applications for the next generation of nano electronics. The electronic properties of graphene, at low 
energy, are governed by the Dirac nature of it's quasi-particle which stems from it's hexagonal lattice geometry~\cite{neto}.
Graphene-like two-dimensional ($2$D) lattice structure with an additional atom at the center of the hexagon can 
be realized in a $\mathcal{T}_3$ or dice lattice~\cite{vidal}. In contrast to graphene, the quasiparticles in this Dirac-Weyl 
material exhibit higher pseudospin $S=1$ states~\cite{vidal}. Also, the presence of this additional atom at the 
center of the hexagon in the dice lattice gives rise to a dispersionless flat band at each Dirac point in addition to 
the Dirac cone found in graphene~\cite{vidal}. In recent times, this type of Dirac-Weyl materials with higher spin 
states, $S=1$, $3/2$, $2$, etc., is attracting much attention aimed at exploring the role of this additional 
atom in them~\cite{malcolm}. A series of investigations has been carried out in order to reveal different 
aspects of the $\mathcal{T}_3$ lattice~\cite{janik,lan,morigi}.

A smooth transition from pseudospin $S=1/2$ (graphene) to $S=1$ (dice or $\mathcal{T}_3$ lattice) can be realized using the $\alpha$-$\mathcal{T}_3$ 
model. Here, $\alpha$ is associated with the strength of the coupling of the central atom to its nearest neighbors. It has 
recently been demonstrated in Hg$_{1-x}$Cd$_{x}$Te that under suitable doping concentration, this material can be 
mapped to $\alpha$-$\mathcal{T}_3$ model~\cite{malcolm} with $\alpha=1/\sqrt{3}$. Moreover, the continuous evolution of $\alpha$ from 
$0$ (graphene) to $1$ ($\mathcal{T}_3$) can be linked to a variable Berry phase by suitably parametrizing $\alpha$~\cite{morigi}. The Berry 
phase, the geometrical phase arising during an adiabatic cyclic evolution of a quantum state, plays a vital role in explaining 
different properties of a system~\cite{Dxiao} such as the dc Hall conductivity~\cite{nicol1}, magneto-transport properties in 
the presence of randomly scattered charged impurities~\cite{tutul}, and optical~\cite{malcolm,nicol3,plasmon} properties. Note that, 
Berry phase is $\pi$ and $0$, respectively in graphene and the $\mathcal{T}_3$ lattice setting the two limits of $\alpha$-$\mathcal{T}_3$ system. 

Magnetotransport measurements have always been appreciated for providing an efficient way to probe a $2$D 
fermionic system. The presence of a magnetic field  perpendicular to the plane of the system drastically changes 
the electronic band structure by the formation of discrete energy levels \ie Landau levels. Fluctuation of chemical 
potential between different Landau levels with respect to the magnetic field manifests itself through the 
appearance of the well known Shubnikov-de Hass oscillation in the longitudinal components ($\sigma_{xx/yy}$) 
of electrical conductivity~\cite{duan}. On the other hand, off-diagonal components ($\sigma_{xy/yx}$) of electrical 
conductivity tensor \ie quantum Hall conductivity become quantized due to the incomplete cyclotron orbits of 
the electrons along the two opposite edges of the $2$D system, transverse to the applied electric field~\cite{duan}.

In the presence of a perpendicular magnetic field, electrons do not posses any finite drift velocity inside the bulk 
of a $2$D system. However, they may acquire a finite drift velocity if the system is subjected to an external 
perturbation. A magnetic field dependent oscillatory drift velocity can be imparted to electrons by applying 
an external perturbation which is periodic in space. This oscillatory drift velocity induces a new type of 
quantum oscillation in the magneto-resistance signal at a low range of magnetic field. This oscillation, known as 
Weiss oscillation, was first observed in magneto-resistance measurements in the electrically modulated 
usual $2$D electronic systems~\cite{Weiss,Gerh,kotha}.  The Weiss oscillation, also known as Commensurability 
oscillation is caused by the commensurability of the two length scales \ie radius of the cyclotron orbit near the 
Fermi energy and the period of the modulation~\cite{poulo,chao,vasilo}. Beenakker \etal~\cite{beenakker} explained 
this oscillation using the concept of {\it guiding-center-drift resonance} between the periodic cyclotron orbit 
motion and the oscillating drift of the orbit center induced by the potential grating. From the application perspective, 
a modulated electric potential may result in a transition from semiconducting to semi-metallic behavior of graphene ~\cite{ho}. 
Under an appropriate class of experimentally feasible one-dimensional external periodic potentials, electron beam 
supercollimation can be realized in graphene as suggested by Park \etal ~\cite{park}.

Apart from electric modulation scenario, magnetic modulation has also been considered 
theoretically~\cite{peeters,super,zhong,matuli,russia,xia,papp}, followed by several experiments~\cite{mag_exp1,mag_exp2,mag_exp3}.
In parallel, a wide range of applications has also been suggested by several groups. In 2007 Chieh \etal have developed a magnetic-fluid 
optical fiber modulators via magnetic modulation~\cite{chieh}. Even a $2$D electron gas ($2$DEG) under spatially modulated magnetic 
field can be used as a fantastic test bed to study resistivity induced by electron-electron scattering~\cite{kato}. There are some other 
applications too~\cite{mewes}. 

Recently, the beating pattern in Weiss oscillation in Rashba spin-orbit coupled electrically/magnetically modulated $2$DEG was 
investigated~\cite{beat_vasi,beat_firoz}. However, Matulis \etal have shown that Weiss oscillation can be enhanced in graphene 
due to the higher Fermi velocity associated with it's linear massless energy dispersion~\cite{matulis_gra}. Tahir \etal have 
studied the same but with magnetic modulation and predicted an enhancement of the amplitude and opposite phase in comparison 
to the case of electrically modulated graphene~\cite{tahir_gra} . Similar investigation have been carried out in electrically 
modulated bilayer graphene~\cite{zarenia} and silicene~\cite{firoz_sili,vasi_sili}. However, modulation induced Weiss 
oscillation has not been addressed to date in the $\alpha$-$\mathcal{T}_3$ system to the best of our knowledge. 

In this article, we investigate the behavior of Weiss oscillation aiming to explore the role of the variable Berry phase in it. Most 
interestingly, we find that the Berry phase causes a valley polarization in magnetoconductivity which is in contrast to that in 
graphene and usual 2DEG. Note that valley polarized magnetoconductivity has been predicted in silicene too, but that is in the presence 
of spin-orbit interaction and a gate voltage\cite{vasi_sili}. However, when both types of modulations are considered together, 
the particle-hole symmetry of the system is broken, leading to an asymmetric density of states (DOS) in addition to the valley polarization in 
magnetoconductivity. We also notice 
\begin{figure}[!thpb]
\centering
\includegraphics[height=4.0cm,width=0.80 \linewidth]{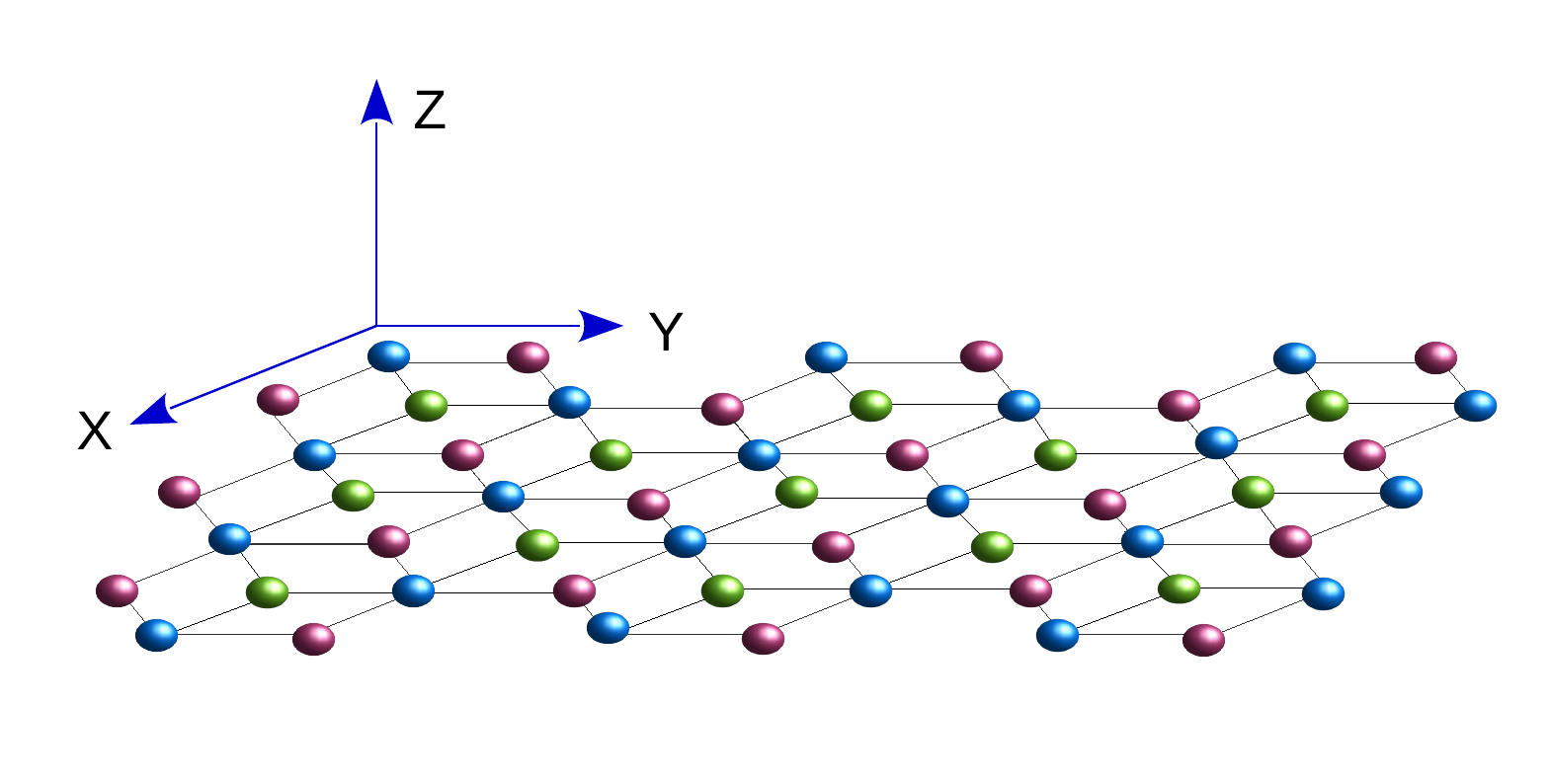}
\caption{(Color online) Schematic of the $\alpha$-$\mathcal{T}_3$ lattice is presented. Colors denote three sub-lattices, 
\ie A (magenta), B (blue) and C (green).}
\label{model}
\end{figure}
a transition from odd (even) to even (odd) filling fraction corresponding to density of states peaks with the variation of the Berry 
phase.

Valley-polarization in transport co-efficients of the $\alpha$-$\mathcal{T}_3$ lattice may have a possible application in valleytronics, a technology 
with control over the valley degree of freedom of the carriers~\cite{qi2014mechanically,ezawa2013spin,gunawan,xiao}. Several proposals 
have been made for valleytronics devices that can be used in encoding or processing information~\cite{schaibley}, and also as 
valley filters or valley valves~\cite{rycerz2007valley}. Intensive research on engineering potential valleytronic devices based on manipulating 
the valley nondegeneracy is needed in order to enrich this newly growing field more and more. 

The remainder of this paper is organized as follows. In Sec.~\ref{sec2}, we describe our model Hamiltonian and the corresponding Landau 
levels. Our results on the effect of weak spatial modulation (both electric and magnetic cases separately) on energy levels, diffusive 
conductivity and valley-polarization are discussed in Sec.~\ref{sec3}. In Sec.~\ref{sec4} we discuss the combined effects of electric 
and magnetic modulation. Finally, we summarize our results and conclude in Sec.~\ref{sec5}.

\section{Model Hamiltonian and formation of Landau levels}\label{sec2}
In Fig.~\ref{model} we present a schematic of an $\alpha$-$\mathcal{T}_3$ lattice which consists of three atoms per unit cell, namely, A, B, 
and C. Atoms A and B form a honeycomb lattice structure similar to graphene with nearest-neighbor hopping amplitude $t$. The 
presence of  an additional third atom C makes this lattice behave differently than graphene. In particular, atom C is connected to
atom B with hopping amplitude $\alpha t$ ($\alpha < 1$).  

Considering the three basis corresponding to the three atoms in the unit cell, the low energy Hamiltonian of this system close to the 
Dirac points around a particular valley can be written as~\cite{tight}
\begin{equation}
\mathcal{H}_0=\left[\begin{array}[c]{ccc}
            0      &  f_{\bf p}\cos\phi &   0 \\
            f_{\bf p}^\ast\cos\phi  &  0  &  f_{\bf p}\sin\phi \\
            0     &  f_{\bf p}^\ast\sin\phi  &  0\end{array}\right].
\end{equation}
Here, $f_{\bf p}=v_F(\eta p_x-ip_y)$ where $\eta=\pm$ denotes the two valleys $K$ and $K^{\prime}$ respectively, 
${\bf p}=\{p_x,p_y\}$ is the $2$D momentum vector and $v_F$ is the Fermi velocity. Note that, the angle $\phi$ is parametrized by 
$\alpha$ via $\alpha=\tan\phi$ and the Hamiltonian is rescaled accordingly. The energy dispersion 
of the conic band can be readily obtained as $E_{k,\lambda}=\lambda \hbar v_{F}k$, with $\lambda=\pm$ correspond 
\begin{figure}[!thpb]
\centering
\includegraphics[height=4.8cm,width=0.58 \linewidth]{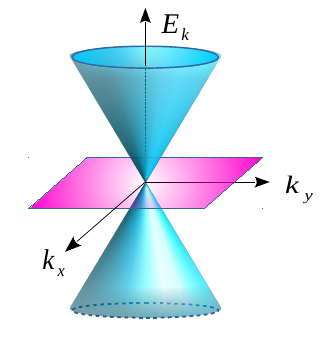}
\caption{(Color online) Schematic of the band structure of $\rm \alpha$-$\mathcal{T}_3$ lattice. The blue area 
represents the conic band structure, while the pink area denotes the dispersionless flat band.}
\label{band}
\end{figure}
to the conduction and valence band, respectively. In addition to the conic bands there is a flat band with $E_{k,0}=0~~\forall~k$. 

The full band structure of the $\alpha$-$\mathcal{T}_3$ model is shown in Fig.~\ref{band}. The eigen states of the conic band are given by
\begin{equation}
\psi_{\lambda}=\left[\begin{array}[c]{c}\cos\phi ~~e^{i\theta_k} \\ \lambda \\ \sin\phi ~~e^{-i\theta_k}
                                \end{array}\right]
\end{equation}
and for the flat band the same can be written as,
\begin{equation}
\psi_0=\left[\begin{array}[c]{c}\sin\phi ~~e^{i\theta_k} \\ 0 \\ -\cos\phi~~ e^{-i\theta_k}
                                \end{array}\right],
\end{equation}
where $\theta_k=\tan^{-1}(k_y/k_x)$. The angle $\phi$ is connected to the Berry phase ($\Omega_{\eta}$) in the conic band 
via $\Omega_{\eta}=\pi\eta\cos(2\phi)=\eta\pi(1-\alpha^2)/(1+\alpha^2)$. In contrast to graphene or the dice lattice the 
Berry phases in this lattice model corresponding to the two valleys are different from each other~\cite{tight}.

\begin{figure}[!thpb]
\centering
\includegraphics[height=7cm,width=0.8 \linewidth]{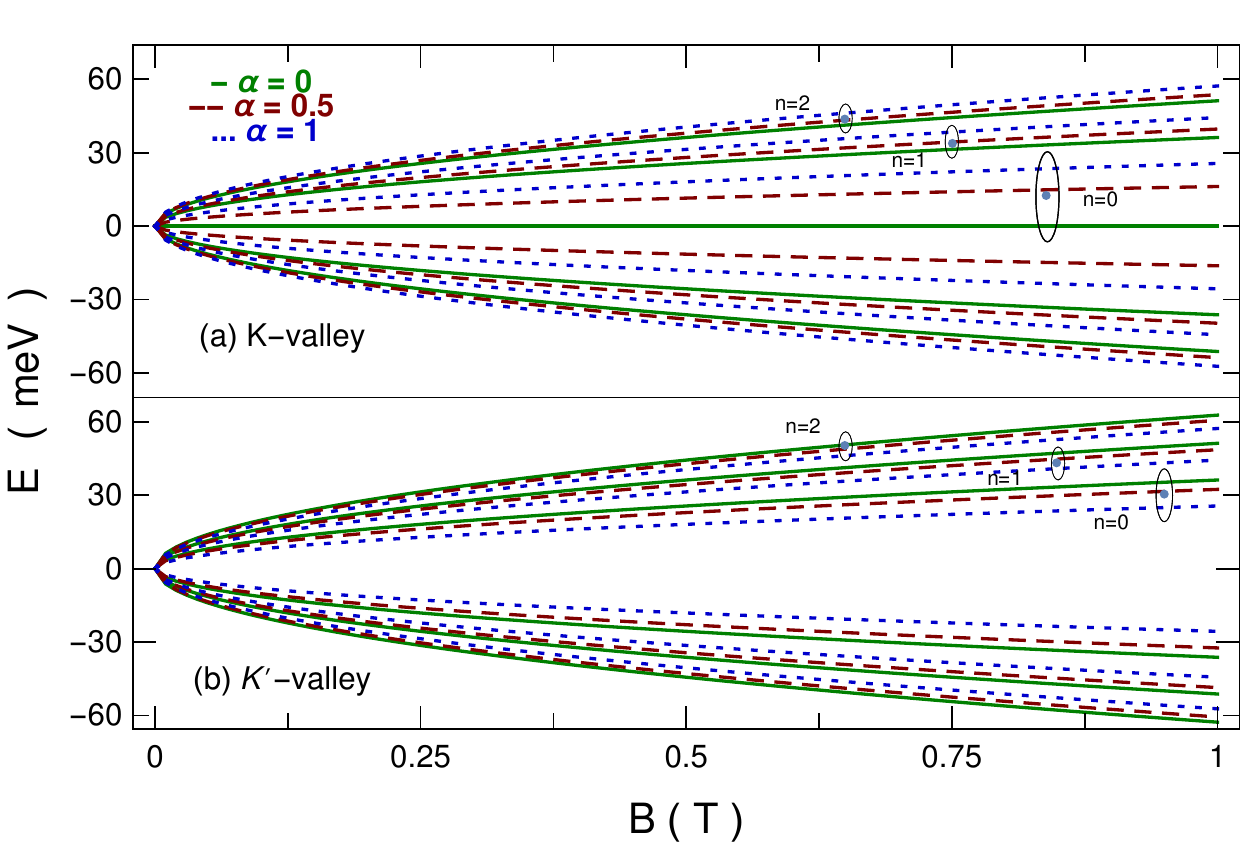}
\caption{(Color online) Plot of a few Landau levels of the $\alpha$-$\mathcal{T}_3$ lattice as a function of the magnetic field 
in the (a) $K$ and (b) $K^{\prime}$ valleys for different values of $\alpha$ is presented.}
\label{landau}
\end{figure}
\begin{figure}[!thpb]
\centering
\includegraphics[height=5.2cm,width=0.7 \linewidth]{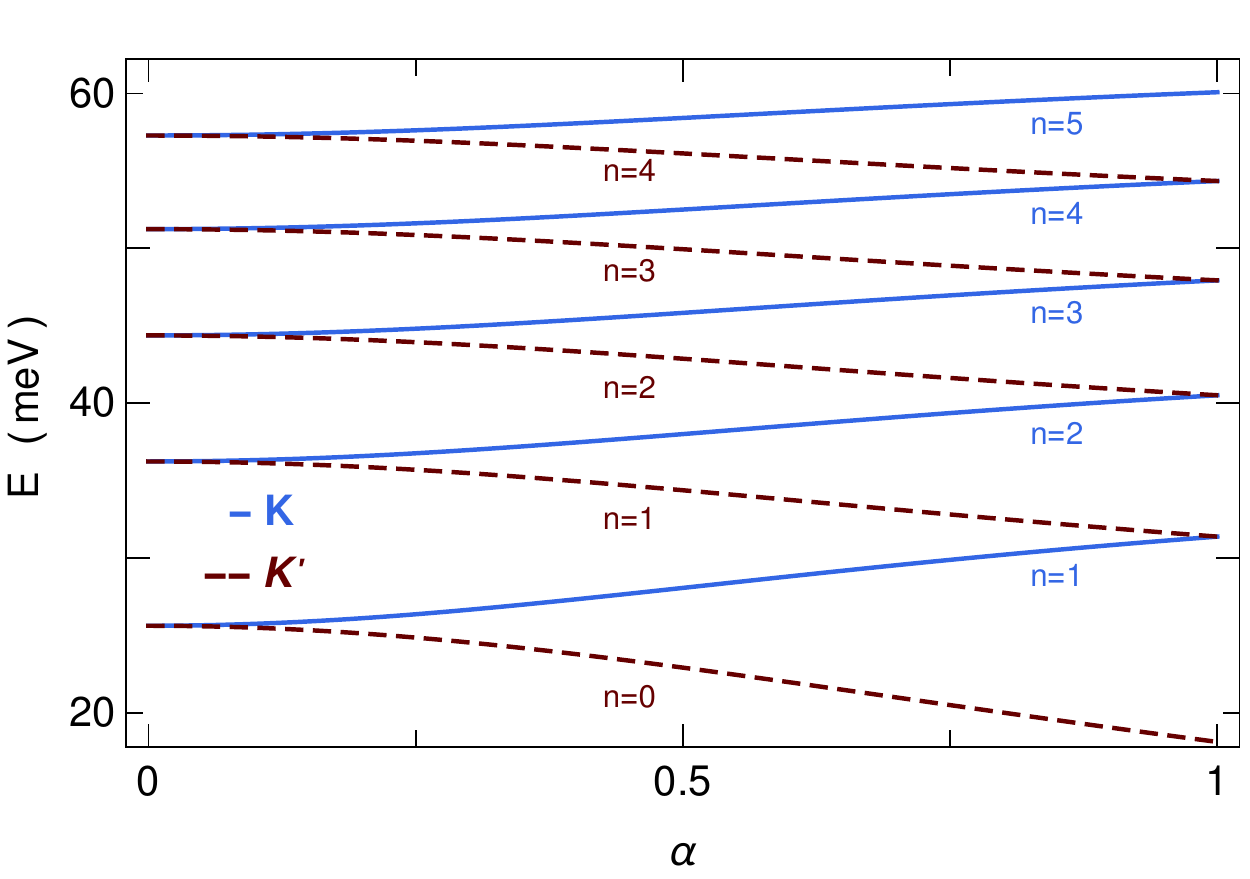}
\caption{(Color online) Plot of a first few Landau levels of the $\alpha$-$\mathcal{T}_3$ lattice as a function of $\alpha$ 
in $K$ (upper-panel) and $K^{\prime}$ (lower panel) valleys at $B=0.5$T.}
\label{Landau_alpha}
\end{figure}
The application of a uniform magnetic field normal to the lattice plane (X-Y) can be incorporated via Peierls 
substitution ${ \Pi}=({\bf p}+e{\bf A}$), where the vector potential ${\bf A}$ is considered under Landau gauge 
as ${\bf A}=(-By,0,0)$ describing the magnetic field ${\bf B}=B\hat{z}$. Hence, the Hamiltonian near the 
Dirac point in the ${K}$ valley reduces to
\begin{equation}\label{HamK}
\mathcal{H}_K=\epsilon
\left[\begin{array}[c]{ccc}
    0      &  \hat{a}\cos\phi\, &   0 \\
    \hat{a}^\dagger\cos\phi\,  &  0  &  \hat{a}\sin\phi\, \\
    0     &  \hat{a}^\dagger\sin\phi\,  &  0
\end{array}\right]
\end{equation}
where $\epsilon=\hbar\omega_c$ with $\omega_c=\sqrt{2} v_F/l_c$. Here $l_c=\sqrt{\hbar/(eB)}$ 
is the magnetic length. Also, $\hat{a}=v_F\Pi_-/\epsilon$ and $\hat{a}^\dagger=v_F \Pi_+/\epsilon$ 
are the usual harmonic oscillator annihilation and creation operators, respectively with $\Pi_{\pm}=\Pi_x\pm i\Pi_{y}$. 
The Hamiltonian for the $K^{\prime}$ valley can be obtained through the substitution 
$\hat{a}\rightarrow -\hat{a}^\dagger$.
%
Hence, diagonalizing Eq.(\ref{HamK}), one can directly obtain the Landau levels of the system in the form~\cite{tight}
\begin{eqnarray}
\label{Energy}
E_{\xi}^\lambda=\lambda \, \epsilon\sqrt{n+\chi_\eta},
\end{eqnarray}
where $\xi\equiv\{n,\eta\}$ corresponds to a set of quantum numbers with $n=0,1,2,...$ being the Landau level index.
The quantity $\chi_\eta$ is related to $\phi$ via the relation,
\begin{equation}
\chi_\eta=[1-\eta\cos(2\phi)]/2.
\end{equation}

In Fig.~\ref{landau}, we show the nature of first few Landau levels ($n=0$, $1$ and $2$) as a function of the magnetic field 
for three different values of $\alpha$. There is a zero energy Landau level for $n=0$ in the $K$ valley. On the other hand, the $n=0$
level is of parabolic form in the $K^{\prime}$-valley. Note that, in graphene both valleys exhibit $n=0$ Landau level. The parabolic 
form can also be induced by tuning the Berry phase from zero to a finite value keeping the valley unaltered. The rest of the 
energy levels are shifted while the Berry phase is tuned. The Berry phase induced energy shift is positive for the $K$ valley. While it 
is negative for the K$^\prime$-valley. 

In Fig.~\ref{Landau_alpha}, we show the variation of the first few Landau levels as a function of the Berry phase. It is evident
from Fig.~\ref{Landau_alpha} that $n$-th Landau level of $K$-valley merges with the $(n+1)$-th Landau level of 
$K^{\prime}$-valley for $\alpha=0$. The evolution of Landau levels was pointed out in Refs.~[\onlinecite{nicol3,nicol1}].

The eigenfunction for $n>0$ corresponding to the $K$ valley is given by
\begin{eqnarray}
\label{ef}
\Psi_{n,k_x}^{\lambda,+}({\bf r})=\frac{e^{ik_xx}}{\sqrt{2L_x}}\left(
\begin{array}{c}
 A^{+}\Phi_{n-1}\big[\frac{y-y_0}{l_c}\big]\\
\lambda \Phi_n\big[\frac{y-y_0}{l_c}\big]\\
B^{+}\Phi_{n+1}\big[\frac{y-y_0}{l_c}\big]
\end{array}\right).
\end{eqnarray}

Here, the cyclotron orbit is centered at $y=y_0=l_c^2k_x$ and
$\Phi_n(y)=[2^nn!l_c\sqrt{\pi}]^{-1/2}\exp(-y^2/2)H_n(y)$ is the usual harmonic oscillator wave function where $H_{n}(y)$ is 
the Hermite polynomial of order $n$. The co-efficients in Eq.~\ref{ef} can be expressed as 
$A^{\eta}=\sqrt{n(1-\chi_{\eta})/(n+\chi_{\eta})}$ and $B^{\eta}=\sqrt{(n+1)\chi_{\eta}/(n+\chi_{\eta})}$.
For the zero-th Landau level \ie $n=0$, the eigenfunction is given by
\begin{eqnarray}
\Psi_{0,k_x}^{\lambda,+}({\bf r})=\frac{e^{ik_xx}}{\sqrt{2L_x}}\left(
\begin{array}{c}
0\\
\lambda\Phi_0\big[\frac{y-y_0}{l_c}\big]\\
\Phi_1\big[\frac{y-y_0}{l_c}\big]
\end{array}\right).
\end{eqnarray}
The wave function in the $K^{\prime}$-valley can be obtained by performing the following transformations
\begin{equation}
\Psi_{n,k_x}^{\lambda,-}=\Psi_{n,k_x}^{\lambda,+}[A^{+(-)}\rightarrow B^{-(+)},
\lambda\rightarrow-\lambda,\Phi_{n-1} \leftrightarrows \Phi_{n+1}].
\end{equation}
Note that, unlike monolayer graphene, the $\rm \alpha$-$\mathcal{T}_3$ lattice exhibits a dispersionless flat band too which is well described
in Ref.~[\onlinecite{tutul}]. As we are considering a doped $\rm \alpha$-$\mathcal{T}_3$ lattice where the Fermi level is well inside the 
conic band, we ignore the effects of the flat band in our analysis.

\section{Effect of electric or magnetic modulation}\label{sec3}
\subsection{Energy correction due to modulation}
We assume in our analysis that the strength of the spatial electric/magnetic modulation is weak compared to 
the Landau level energy scale such that we can treat the modulation perturbatively. The first
order energy correction is evaluated in all cases by using the unperturbed eigen states as follows.
\begin{figure*}
\subfigure[]
{\includegraphics[width=.49\textwidth,height=5cm]{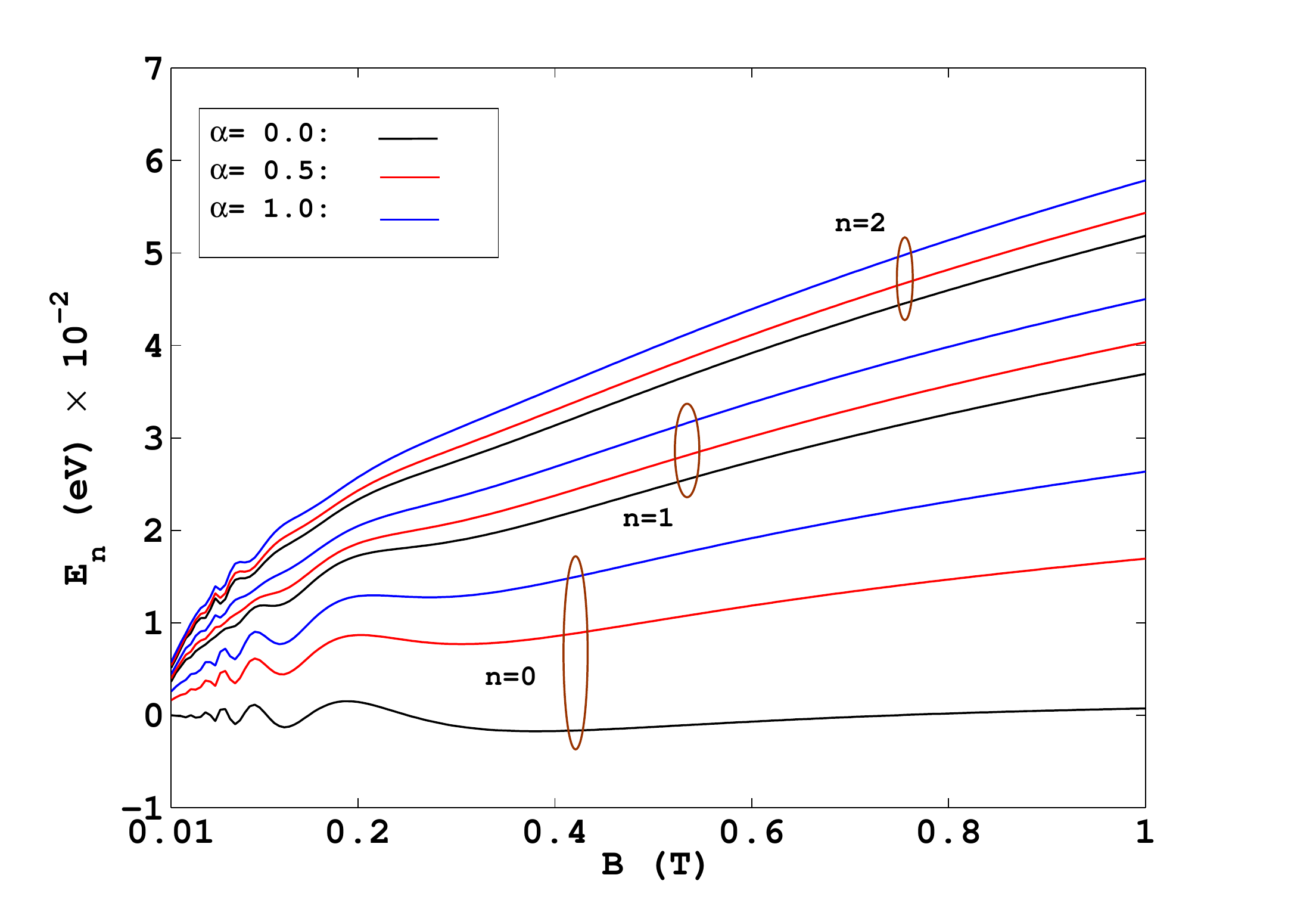}}
\subfigure[]
{\includegraphics[width=.49\textwidth,height=5cm]{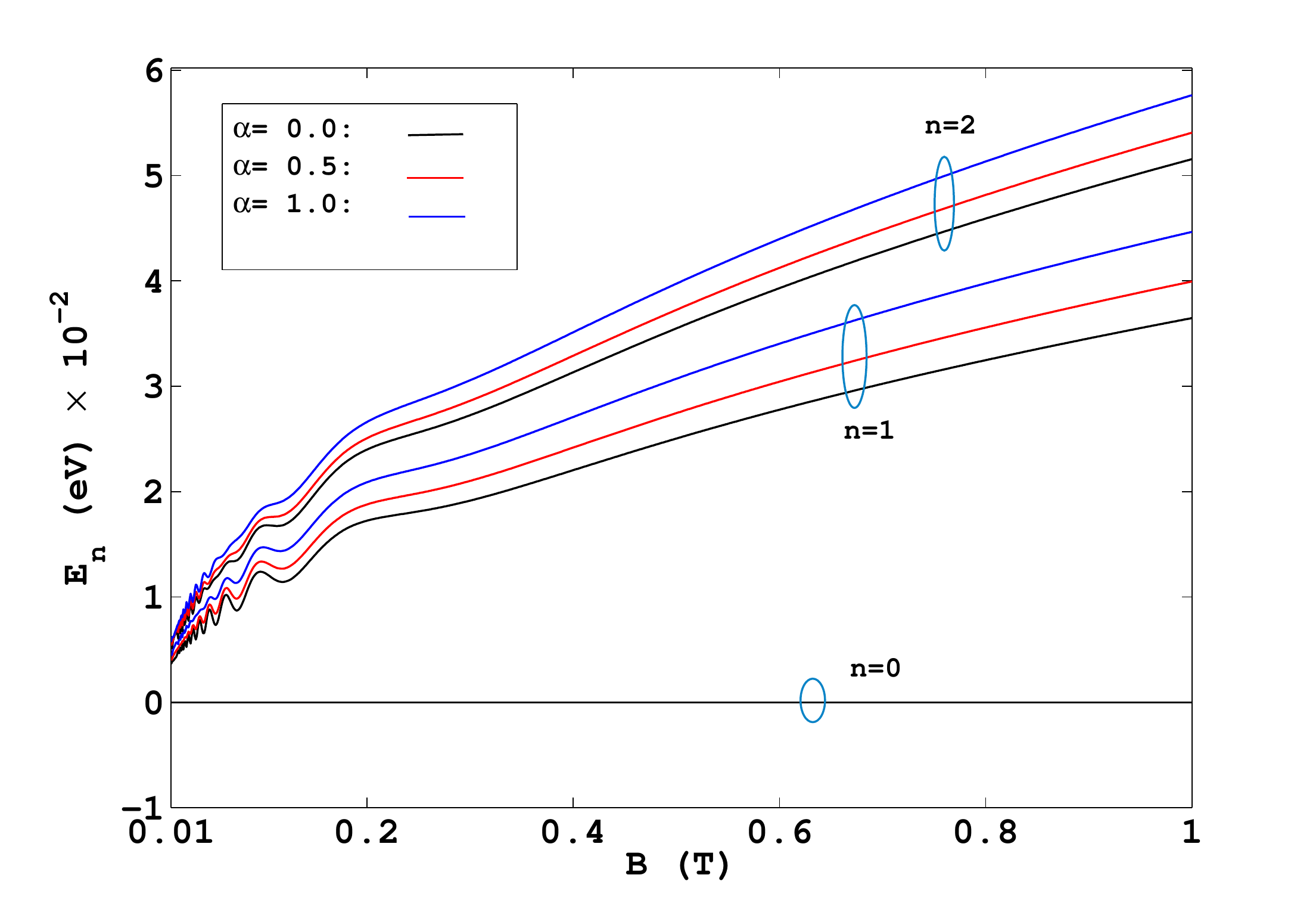}}
\caption{(Color online) First few modulated Landau levels in $K$ valleys. (a) Electrically modulated and (b) magnetically modulated 
Landau levels. Here, the parameters are chosen to be modulation strength $V_e=V_m=1~$meV, $k_x=10^8~m^{-1}$ and modulation period 
$a=350$ nm.} \label{modulated_LL}
\end{figure*}
\subsubsection{Electric modulation}
We describe our electrically modulated system by the Hamiltonian $\mathcal{H}_T^e=\mathcal{H}_{\eta}+V_{e}\cos(q y)$, where 
$V_e$ is the strength of the electric modulation and $q=2\pi/a$, where $a$ is the period. Using perturbation 
theory, we evaluate the first order energy correction for the $K$ or $K^{\prime}$-valley as (for $n\ge 1$)
\begin{eqnarray}\label{corr_elec}
\Delta E_{\xi,k_x}^{e}&=&\int_0^{L_x}dx\int_{-\infty}^{\infty}[\Phi_{n,k_x}^{\lambda,\eta}({\bf r})]^{\ast}V_{e}\cos(q y)
\Phi_{n,k_x}^{\lambda,\eta}({\bf r})dy\nonumber\\
 &=&\frac{V_{e}}{2}F_{\xi}(u)\cos(q y_0).
\end{eqnarray}
 Here 
\begin{equation}\label{Fexp}
F_{\xi}(u)=e^{-\frac{u}{2}}[\mid A^{\eta}\mid^2 L_{n-1}(u)+L_{n}(u)+\mid B^{\eta}\mid^2 L_{n+1}(u)]
\end{equation}
where $L_{n}(u)$ is the Laguerre polynomial of order $n$ and $u=q^2l_c^2/2$. The total energy is now 
$E_{\xi,k_x}^{e}=E_{\xi}+\Delta E_{\xi,k_x}^{e}$ where $k_x$ degeneracy is lifted. The energy correction to the ground state 
($n=0$) is
\begin{equation}
\Delta E_{\{0,+\},k_x}^{e}=\frac{V_e}{2}e^{-\frac{u}{2}}[L_{0}(u)+L_{1}(u)]\cos(q y_0).
\end{equation}

In Fig.~\ref{modulated_LL}(a), the features of the first few modulated Landau levels of the $K$ valley are shown as a function of 
the magnetic field for different values of $\alpha$. For lower values of magnetic field, electrical modulation induces a 
sinusoidal nature to the Landau level and this feature slowly disappear with the increase of Landau level index $n$. The 
qualitative behavior of the modulated energy levels in $K^{\prime}$-valley is similar to that in $K$ valley. 

\subsubsection{Magnetic modulation}
Now we consider the case where the perpendicular magnetic field is weakly modulated without any electrical modulation. The dynamics 
of charge carriers under modulated magnetic field is believed to be closely related to {\it composite fermions} in the fractional 
quantum Hall regime\cite{composite_fermion}. Under a weak magnetic field regime, theoretical works exist from the usual $2$DEG to monolayer 
graphene (mentioned in introduction) and extracted modulation induced Weiss contribution. Along the same line, we investigate Weiss 
oscillation in a magnetically modulated $\alpha$-$T_3$ lattice.

First, we evaluate the first order energy correction due to magnetic modulation. Let the perpendicular magnetic field be modulated 
very weakly as ${\bf B}=[B+B_{m}\cos(q y)]\hat{z}$, where $B_{m}\ll B$ describes the vector potential under the Landau gauge 
${\bf A}=[-By-(B_m/q)\sin(q y),0,0]$. Similarly to the case of electric modulation, the total Hamiltonian can now be splitted 
into two parts as $\mathcal{H}_T^m=\mathcal{H}_{\eta}+\mathcal{H}_m$, where $\mathcal{H}_{\eta}$ is the unperturbed Hamiltonian
and $\mathcal{H}_{m}$ is the modulation induced 
perturbation which can be written as
\begin{equation}
 \mathcal{H}_{m}=\left[\begin{array}[c]{ccc}0 & \Upsilon\cos\phi & 0\\
        \Upsilon\cos\phi& 0& \Upsilon\sin\phi\\
        0& \Upsilon\sin\phi& 0\\
        \end{array}\right],
\end{equation}
where $\Upsilon= \eta eB_m v_{F}\sin(q y)/q$. Using the unperturbed wave function, the first order energy correction due to magnetic 
modulation $H_{m}$ can be evaluated as (for $n\ge 1$)
\begin{equation}\label{corr_mag}
\Delta E_{\xi,k_x}^{m}=\frac{V_m}{2} G^{\lambda}_{\xi}(u)\cos(q y_0)
\end{equation}
with $G^{\lambda}_{\xi}(u)=\lambda\eta R_{\{n,\eta\}}(u)$. For the $K$ valley,
\begin{eqnarray}\label{Rexp}
R_{\{n,+\}}(u)=\frac{2}{ql_c} e^{-\frac{u}{2}}[A^{+}\Lambda_{n}(u)\cos\phi+B^{+}\Lambda_{n+1}(u)\sin\phi] \nonumber \\
\end{eqnarray}
with $\Lambda_{n}(u)=\sqrt{2n}\{L_{n-1}(u)-L_{n}(u)\}$ and $V_m=\hbar\omega_m$ where $\omega_{m}=ev_{F}B_{m}/q$. Similarly, 
one can obtain the first order energy correction in the $K^{\prime}$ valley by interchanging the sine and cosine terms in $R_{\{n,-\}}(u)$,
and $\lambda\rightarrow-\lambda$. Akin to the electric modulation case, here also we observe the oscillation 
of first few Landau levels due to the effect of magnetic modulation for the same strength of modulation ($V_e=V_m=1$ meV) as displayed in 
Fig.~\ref{modulated_LL}(b). Note that, modulation does not have any effect on the first order energy correction of the $n=0$ level.
This is shown by the straight horizontal line. In the $K^\prime$ valley, the behavior of the energy correction for different Landau levels
is similar in nature qualitatively.

\subsection{Diffusive conductivity}

Under a low temperature regime there are mainly two contributions to the longitudinal conductivity, one is the impurity induced 
collisional conductivity and the other one is the modulation induced diffusive conductivity. However, the dominant contribution 
to the conductivity under a low magnetic field arises from the electron diffusion, caused by the applied in-plane weak spatial 
modulation. So, throughout the rest of the article, we use the term diffusive conductivity to refer to magnetoconductivity. 
On the other hand, the quantum Hall conductivity corresponding to the off-diagonal component of the conductivity tensor is neglected
here due to the minor effect of the modulation as revealed in the literature~\cite{vasilo}. In order to compute the diffusive 
contribution to the longitudinal conductivity, \ie,
\begin{figure*}
 \subfigure[]
 {\includegraphics[width=.49\textwidth,height=6cm]{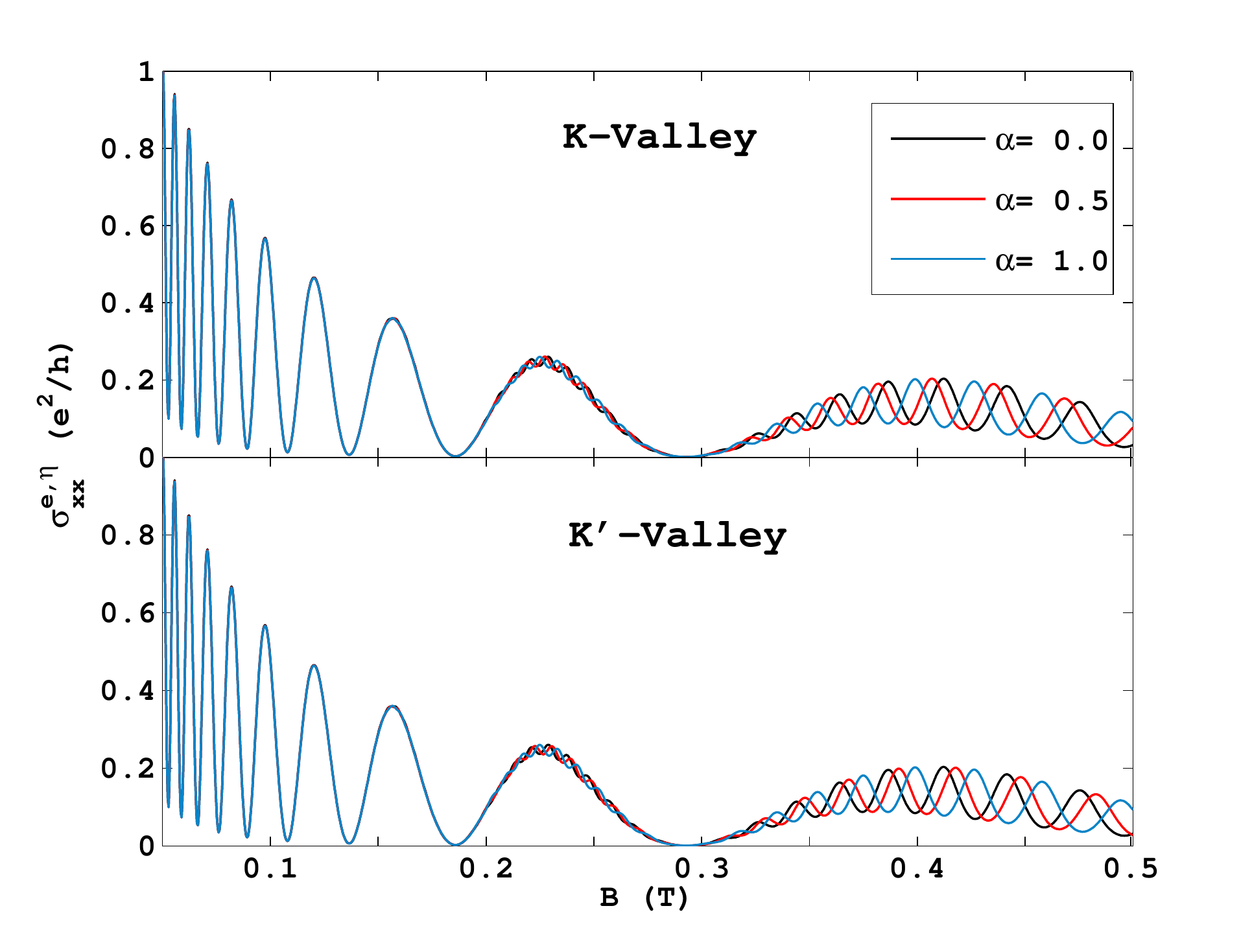}}
 \subfigure[]
  {\includegraphics[width=.49\textwidth,height=6cm]{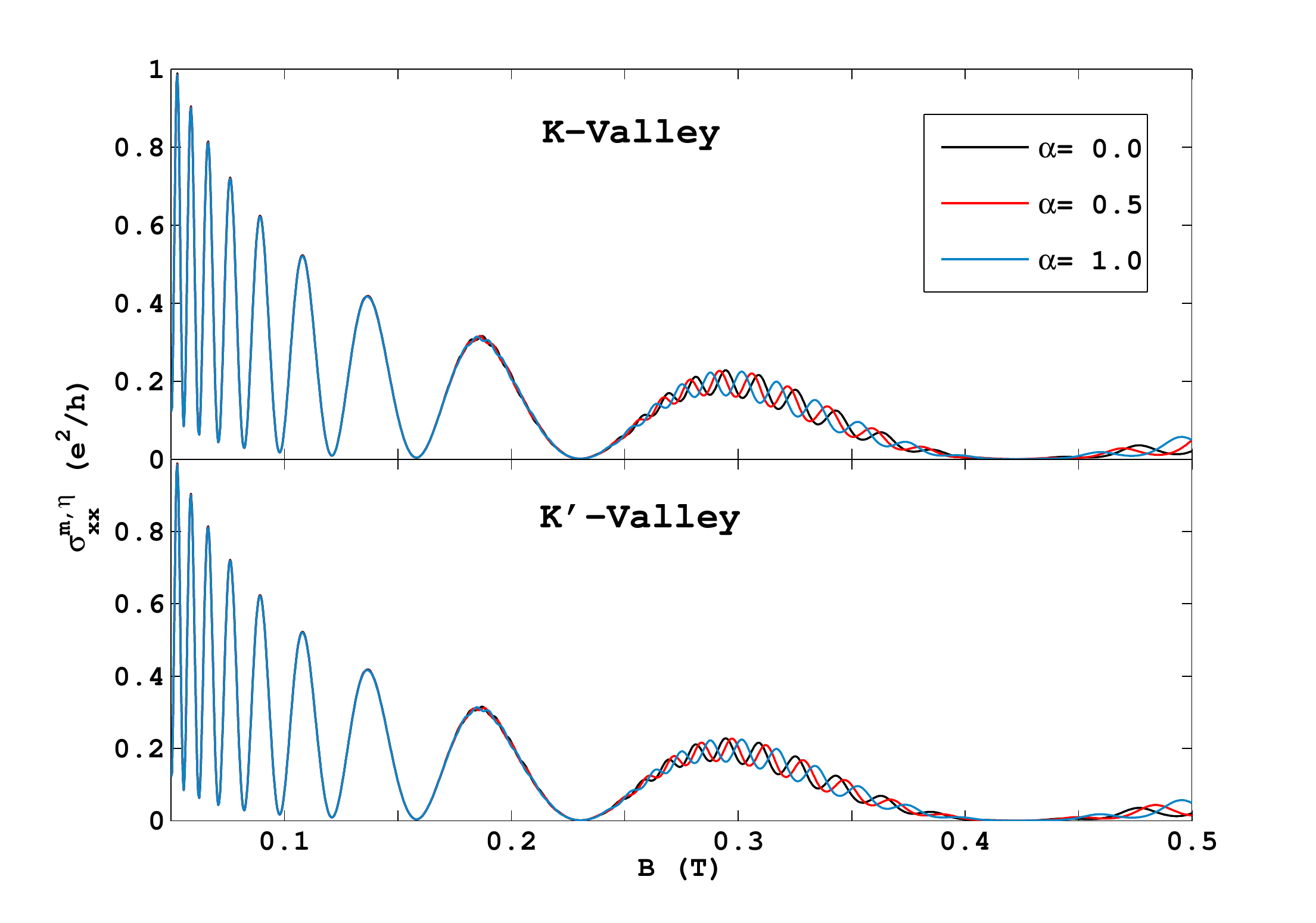}}
 \caption{(Color online) Weiss oscillation for (a) electric modulation and (b) magnetic modulation. Variation of diffusive conductivity 
 is displayed as a function of the perpendicular magnetic field for the $K$ (upper panel) and $K^{\prime}$ (lower panel) valley 
 choosing three different values of $\alpha$.}
 \label{diff_con}
 \end{figure*}
diffusive conductivity, we adopt the formalism developed in Ref.~[\onlinecite{vilet}], which has been employed in case of 
$2$DEG~\cite{vasilo} as well as graphene~\cite{matulis_gra,tahir_gra}.

The dc diffusive conductivity can be evaluated by using the following semiclassical expression of the Kubo formula as:~\cite{vilet}
\begin{equation}\label{diff}
\sigma_{\mu\nu}^{\rm }=\frac{\beta e^2}{\Omega}\sum_{\zeta}f_{\zeta}(1-f_{\zeta})\tau(E_\zeta)v_{\mu}v_{\nu}
\end{equation}
provided the scattering processes involved are elastic or quasielastic. Here, $\zeta\equiv\{\xi,k_x\}$, 
$f_{\zeta}=[1+\exp\{\beta(E_{\zeta}-E_{F})\}]^{-1}$ is the Fermi-Dirac distribution function
with $E_F$ is the Fermi energy. In the above formula, $\tau(E_{\zeta})$ denotes the energy dependent collision time
and $v_{\mu(\nu)}=(1/\hbar)\partial E_{\zeta}/\partial k_{\mu(\nu)}$ with $\mu(\nu)=x$ or $y$.
$\Omega=L_{x}\times L_{y}$  is the dimension of the $2$D lattice.

\subsubsection{Electric modulation}
We evaluate the drift velocity $v_{\mu(\nu)}$ in the presence of electric modulation. They are expressed as
\begin{eqnarray}
v^{e}_x&=&\frac{1}{\hbar} \frac{\partial E_{\xi,k_x}^{e}}{\partial k_x}=-\frac{V_e}{\hbar q}uF_{\xi}(u)\sin(q y_0)
\end{eqnarray}
and $v_{y}^{e}=0$. The latter suggests that diffusive conductivity arises along the direction normal to the applied modulation. 
Now we substitute $v_x^{e}$ in Eq.(\ref{diff}) to obtain the diffusive conductivity. We replace the summation over $k_x$ 
with the integral as $\sum_{k_x}\rightarrow\frac{L_x}{2\pi}\int_0^{L_y/l_c^2}dk_x$ by using the fact that the origin of the cyclotron 
orbit is always confined within the system \ie $0\le \mid y_0 \mid \le L_y$. The factor $L_x/(2\pi)$ takes care of the periodic 
boundary condition. Thus the diffusive conductivity simplifies to
\begin{equation}\label{el_diff}
 \sigma_{xx}^{e,\eta}=\frac{e^2}{h}\frac{\beta}{4\Gamma_0} V_e^2 u\sum_{\xi}f_{\xi}(1-f_{\xi})[F_{\xi}(u)]^2,
\end{equation}
where $\Gamma_0$ is the impurity-induced broadening. Here, we assume that the collisional time $\tau(E_{\xi})$ varies very slowly 
with the energy \ie $\tau(E_{\xi})\simeq \tau_{_0}$ which is a valid approximation under a low magnetic field and also substitute 
$\Gamma_0\approx\hbar/\tau_{_0}$. The modulation effect on Fermi distribution function is very small, and hence, we ignore it.

\subsubsection{Magnetic modulation}
Similarly to the case of electric modulation, the magnetic modulation induced drift velocity is given by,
\begin{equation}
v^{m}_x=-\frac{V_m}{\hbar q}uG^{\lambda}_{\xi}(u)\sin(q y_0)
\label{v_mag}
\end{equation}
which leads to the exact form of the diffusive conductivity as
\begin{equation}
\sigma_{xx}^{m,\eta}=\frac{e^2}{h}\frac{\beta}{4\Gamma_0} V_m^2 u\sum_{\xi}f_{\xi}(1-f_{\xi})[R_{\xi}(u)]^2.
\label{sigma_mag}
\end{equation}
In comparison with the electric modulation scenario, the form of the drift velocity in the magnetic modulation case is exactly
the same except for the term containing Laguerre polynomials ($F_{\xi}$(u) and $R_{\xi}(u)$ in the electric and magnetic modulation 
cases, respectively).
\subsubsection{Discussion}
To plot the diffusive conductivity for both the electric and the magnetic modulated systems we choose the following system 
parameters: modulation period $a=350$ nm, temperature $T=6$ K, strength of modulation $V_{e}=V_m=1$ meV and the impurity 
induced Landau level broadening is assumed to be $\Gamma_0=1$ meV. We consider the Fermi energy which corresponds to the carrier 
density $n_e=3\times 10^{15}$ m$^{-2}$.

We show the behavior of the diffusive conductivity as a function of the magnetic field in Fig.~\ref{diff_con} for both the electric
[Fig.~\ref{diff_con}(a)] and magnetic [Fig.~\ref{diff_con}(b)] modulation cases. We observe that under a low magnetic field ($B<0.3$ T), 
the diffusive conductivity exhibits oscillation which is purely of the Weiss type. The variation of $\alpha$ as well as the Berry phase 
does not affect this type of oscillation significantly in this region. However, under the regime of a relatively higher magnetic field 
($B>0.3$ T), the effect of $\alpha$ becomes visible and it causes a phase shift to the Shubnikov-de Haas (SdH) oscillations which are 
superimposed over the Weiss oscillation. Note that, in the $K$ valley the corresponding phase shift advances with an increase in $\alpha$. 
On the other hand, it lags in the $K^{\prime}$ valley. The different Berry phases acquired by the electrons in $K$ and $K^{\prime}$ valleys 
are responsible for the difference in the phase modulation. 

In order to understand the appearance of the valley-dependent phase shift in SdH oscillation over Weiss oscillation, we derive the 
approximate analytical form of the diffusive conductivity in each valley by replacing 
$\sum\limits_{n} \rightarrow \pi l_c^2\int D_{\eta}(E)dE$ in Eq.(\ref{el_diff}). Here, $D_{\eta}(E)$ is the DOS.
Following the approach as discussed in Ref.[\onlinecite{firoz_topo}], we obtain the simplified form of the DOS as
\begin{equation}
 D_{\eta}(E)= D_0 \Big[ 1+2\Omega(E)\cos{\left\{2 \pi \left(\frac{E^2}{\epsilon^2}-\chi_{\eta}\right)\right\}}\Big],
\end{equation}
where $D_0=E/(2\pi \hbar^2 v_{F}^2)$ is the zero magnetic field density of states and impurity induced damping factor
\begin{equation}
 \Omega(E)=\exp\left\{-2 \left(\frac{\pi E \Gamma (E)}{\epsilon^2}\right)\right\}
\end{equation}
with $\Gamma(E)\simeq4\pi \Gamma_0^2 E/\epsilon^2$. After plugging it into Eq.(\ref{el_diff}) and using
the higher Landau level approximation\cite{vasilo} \ie
\begin{equation}\label{asymp}
e^{-u/2}L_{n}(u)\rightarrow \frac{1}{\sqrt{\pi\sqrt{nu}}}\cos\big(2\sqrt{nu}-\frac{\pi}{4}\big)
\end{equation}
we have
\begin{eqnarray}\label{sdhe}
 \sigma_{xx}^{e,\eta}&=&\frac{e^2}{h}\frac{V_e^2}{\Gamma_0}
 \frac{\beta_{W}U_{\eta}}{32\pi^2 }\Big\{\mathcal{W}_e+2\Omega(E_F) R^S\big(\frac{T}{T_S}\big)\nonumber\\&&
 \cos\left[2 \pi k\left(\frac{g}{B} -\chi_{\eta}\right)\right]\cos^2\left[2\pi\left(\frac{f^{\eta}}{B}-\frac{1}{8}\right)-
 \theta^e_{\eta}\right]\Big\}.\nonumber\\
\end{eqnarray}
Here, $\beta_{W}=(k_BT_{W})^{-1}$ with $T_{W}=eav_FB/[4\pi^2 k_B(1-\frac{\chi_{\eta}}{k_{F}^2l_c^2})]$, is the characteristic 
temperature for Weiss oscillation and 
\begin{eqnarray}
U_{\eta}=&&[1+(|A^{\eta}|^2+|B_{\eta}|^2)\cos\nu +(|B^{\eta}|^2-|A^{\eta}|^2)\sin\nu]^2 \nonumber \\
\end{eqnarray}
at $n=n_{F}$ with $\nu=2\pi/(ak_F)$. The appearance of a valley dependent phase factor in the cosine square term is
given by 
\begin{eqnarray}\label{phase}
 \tan\theta^e_{\eta}=\frac{(|B^{\eta}|^2-|A^{\eta}|^2)\sin\nu}{1+(|A^{\eta}|^2+|B_{\eta}|^2)\cos\nu}.
\end{eqnarray}
The first term in Eq.~(\ref{sdhe}), $\mathcal{W}_e$, represents the pure Weiss oscillation, which is given by
\begin{eqnarray}
 \mathcal{W}_e&=& 1-R^W\big(\frac{T}{T_W}\big) \nonumber \\
 &&+2R^W\big(\frac{T}{T_W}\big)
 \cos^2\left[2\pi\left(\frac{f^{\eta}}{B}-\frac{1}{8}\right)-\theta^e_\eta\right].
\end{eqnarray}
The Weiss oscillation frequency $f^{\eta}=[1-\chi_{\eta}(k_{F}^2l_c^2)^{-1}]\hbar k_F/(ea)$, is weakly sensitive to the valley index 
and does not contribute sufficiently to the valley polarization. Also, it exhibits a valley dependent phase factor ($\theta^e_{\eta}$) 
which is too small to make any substantial changes between two valleys because of the small value of $\sin{\nu}$ in the numerator of 
Eq.~(\ref{phase}).

The thermal damping factor describing the decay of the Weiss oscillation amplitude with increasing temperature is given by
\begin{equation}
R^W\big(\frac{T}{T_W}\big)=\frac{T/T_W}{\sinh(T/T_W)}.
\end{equation}
Here, $T_W$ determines the critical temperature beyond which Weiss oscillation starts to die out.

On the other hand, the second term in Eq.~(\ref{sdhe}), containing the product of two cosines, represents the overlapping of SdH oscillation 
over Weiss oscillation with the increase in the magnetic field. The frequency of SdH oscillation ($g=\hbar k_{F}^2/e$) is independent of the 
valley index. The characteristic temperature for SdH oscillation is $T_S=(\hbar\omega_c)^2/(4\pi^2 E_{F})$, with the thermal damping factor 
expressed as
\begin{equation}
R^S\big(\frac{T}{T_S}\big)=\frac{T/T_S}{\sinh(T/T_S)}.
\end{equation}
Similarly to Weiss oscillation, $T_S$ is the critical temperature for the SdH oscillation. The critical temperature for Weiss oscillation 
under a particular magnetic field is higher than that for SdH oscillation.  

From Eq.~(\ref{sdhe}) it is observed that SdH oscillation superimposed over the Weiss region exhibits a valley-dependent phase factor,
$\chi_\eta$, which is the main reason behind the appearance of valley polarization in magnetoconductivity with increasing magnetic field. The 
presence of $U_{\eta}$ in Eq.~(\ref{sdhe}) indicates that the amplitudes of the oscillation in the two valleys are different for the 
entire range of magnetic field, but it's contribution to valley polarization very small. The modulation induced correction to the 
DOS~\cite{firoz_mag_modu} is of the order of $V_{e(m)}^2$ and the corresponding correction to the diffusive conductivity becomes of 
the order of $V_{e(m)}^4$. This is very small and hence we neglect the modulation effect on the DOS. 

For a particular valley, the Berry phase affects only the phase part keeping the amplitude almost the same. This phase shift due to the Berry 
phase enters through the $\chi_{\eta}$ term which is very clear in the higher magnetic field regime. In the context of amplitudes,
$U_{\eta}$ is not too sensitive to the Berry phase to be visualized.

Following the same procedure, we obtain the analytical expression of the diffusive conductivity for magnetic modulation as
 \begin{eqnarray}\label{sdhm}
 \sigma_{xx}^{m,\eta}&=&\frac{e^2}{h}\frac{V_m^2}{\Gamma_0}
 \frac{\beta_{W}M_{\eta}}{32\pi^2}\Big\{\mathcal{W}_m+2\Omega(E_F) R^S\big(\frac{T}{T_S}\big)\nonumber\\&&
 \cos{\left[2 \pi k\left(\frac{g}{B} -\chi_{\eta}\right)\right]}\sin^2\left[2\pi\left(\frac{f^{\eta}}{B}
 -\frac{1}{8}\right)-\theta_{\eta}^m\right]\Big\}.\nonumber\\
\end{eqnarray}
where $\mathcal{W}_m$ represents the pure Weiss oscillation, which is given by
\begin{eqnarray}
 \mathcal{W}_m&=&1-R^W\big(\frac{T}{T_W}\big)\nonumber \\
 &&+2R^W\big(\frac{T}{T_W}\big)\sin^2\left[2\pi\left(\frac{f^{\eta}}{B}-\frac{1}{8}\right)-\theta_{\eta}^m\right].
\end{eqnarray}
and
\begin{eqnarray}
M_{+}&=&[\{(A^{+}\cos\phi+B^{+}\sin\phi)(1-\cos\nu)\}^2\nonumber\\&&+\{(A^{+}\cos\phi-B^{+}\sin\phi)\sin\nu\}^2]^{1/2}.
\end{eqnarray}
The form of $M_{-}$ can be obtained by just replacing $A^{+}\rightarrow B^{-}$ and $B^{+}\rightarrow A^{-}$. The phase factor is 
given by
\begin{equation}
\tan\theta_{+}^{m}=\frac{(A^{+}\cos\phi-B^{+}\sin\phi)\sin\nu}{(A^{+}\cos\phi+B^{+}\sin\phi)(1-\cos\nu)}.
\end{equation}
The expressions for the pure Weiss oscillation for magnetic modulation ($\mathcal{W}_m$) is similar qualitatively to that for electric modulation 
($\mathcal{W}_e$) except they have opposite phases (provided $\theta_{\eta}^e$ and $\theta_{\eta}^m$ are very small)  due to the presence of the
squares of the sine and cosine functions in the two cases, respectively. This phase relationship is similar to that in usual $2$DEG and graphene
which was already pointed out in Refs.~[\onlinecite{peeters}] and [\onlinecite{tahir_gra}].

\begin{figure*}
\subfigure[]
{\includegraphics[width=.49\textwidth,height=5cm]{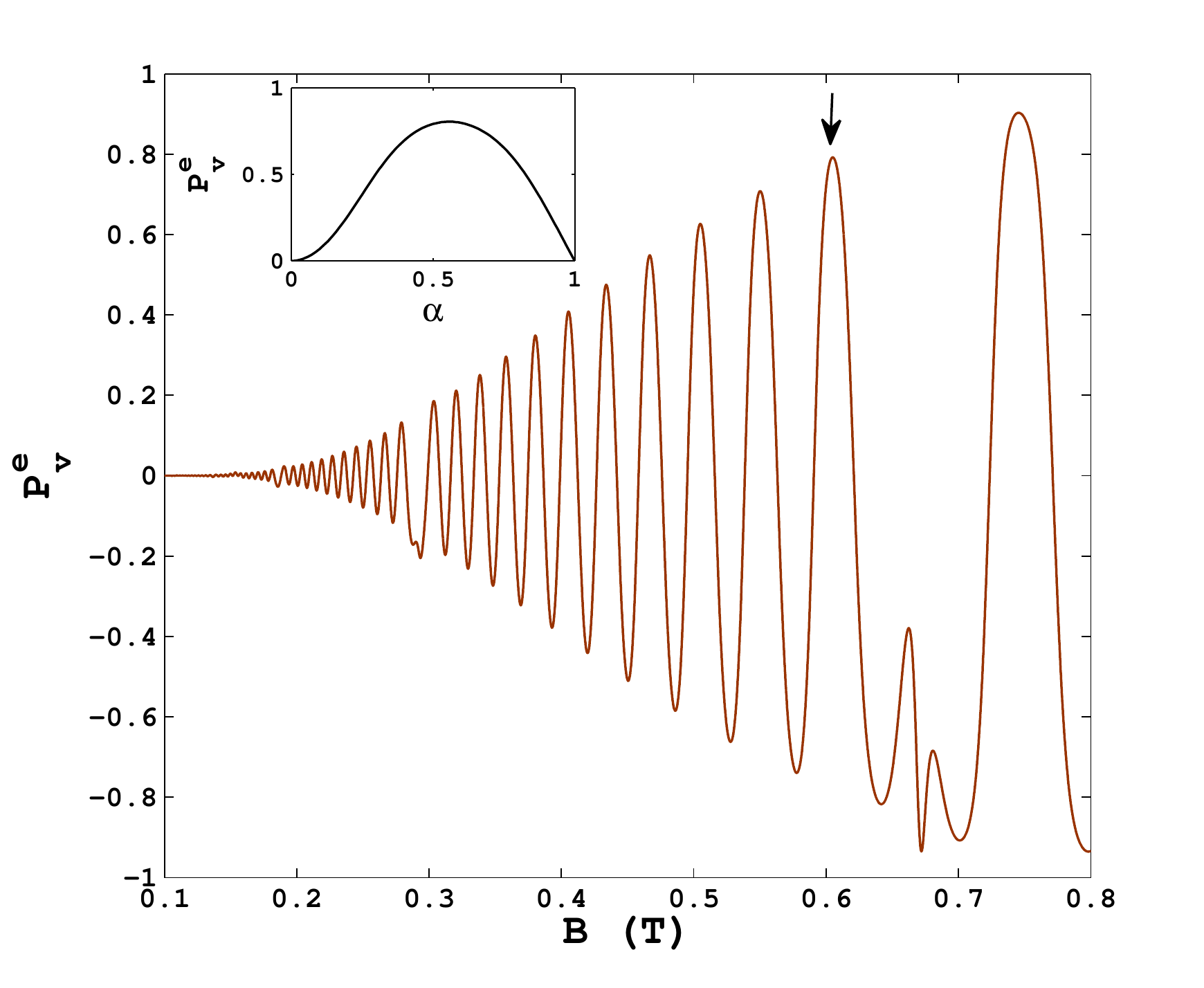}}
\subfigure[]
{\includegraphics[width=.49\textwidth,height=5cm]{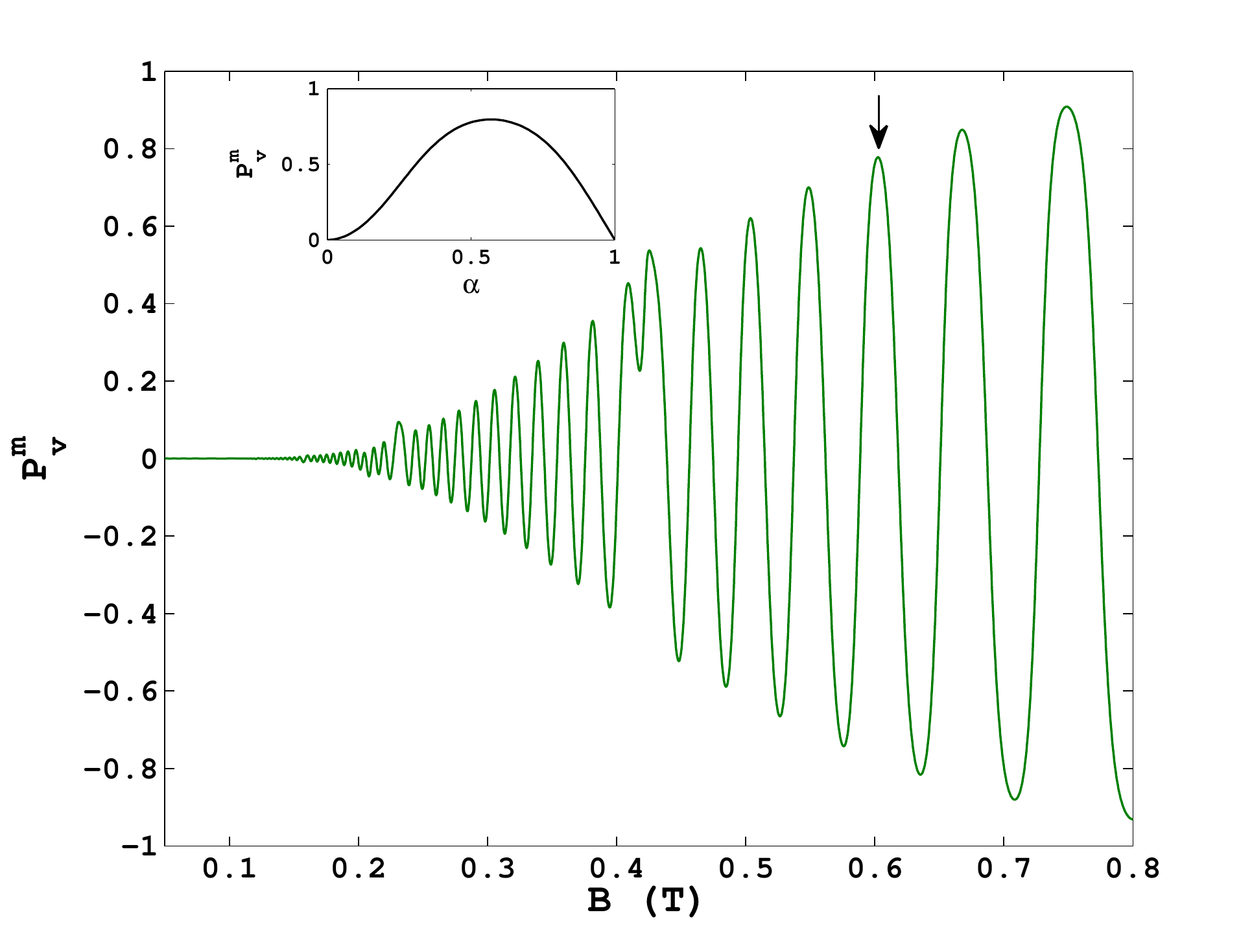}}
\caption{(Color online) Valley polarization for (a) electric modulation and (b) for magnetic modulation is plotted as a function of the magnetic 
field for $\alpha=0.5$. The inset illustrates the features of valley polarization with respect to $\alpha$ corresponding to the magnetic field 
indicated by the black arrow.}
\label{pola}
\end{figure*}
From the above discussion it is clear that the diffusive conductivity is sensitive to the valley degree of freedom. Therefore, it is 
now of interest to examine the valley dependency of the diffusive conductivity in terms of the valley polarization with respect to the 
magnetic field as well as $\alpha$.

\subsection{Valley-polarization}
The valley polarization of diffusive conductivity can be defined as,
\begin{equation}\label{pola_defn}
P_{v}^{\rho}=\frac{\sigma_{xx}^{\rho,{K}}-\sigma_{xx}^{\rho,{K}^{\prime}}}{\sigma_{xx}^{\rho,{K}}+\sigma_{xx}^{\rho,{K}^{\prime}}}
\end{equation}
where $\rho$ may be $\rm e$ or $\rm m$ corresponding to the electric and magnetic modulation case, respectively. We explore the 
valley-polarization due to the finite differences in the diffusive conductivity in the two valleys at $\alpha=0.5$. We calculate the 
valley polarization using Eq.(\ref{pola_defn}) and depict the behavior of valley polarization with respect to the magnetic field in 
Figs.~\ref{pola}(a) and (b) which correspond to the electric and magnetic cases respectively. We observe that under the low magnetic field 
regime where pure Weiss oscillation arises, the valley-polarization is too small to be realized. However, with an enhancement of the 
magnetic field, when SdH oscillation starts to superimpose over Weiss, the polarization increases and rapidly oscillates. It is also possible 
to obtain $90\%$ valley polarization under a high magnetic field regime. We show the result of valley polarization for a particular $\alpha$.
In order to examine the behavior of $P_v$ as a function of the Berry phase we plot $P_v$ vs. $\alpha$ in the insets in Fig.~\ref{pola}.
For this, we fix the magnetic field at the two different values shown by the vertical arrows in Figs.~\ref{pola}(a) and (b) corresponding to the 
electric and magnetic modulation cases, respectively. We notice that valley polarization is maximum for the intermediate $\alpha$ with zeros 
for the two limiting values of $\alpha$ ($0$ and $1$) for both electric and magnetic modulations. For the two limiting values of $\alpha$,
corresponding to graphene ($\alpha=0$) and the dice lattice ($\alpha=1$) the valley degeneracy is recovered and valley polarization disappears. 
Note that, there are two dips in the $P_e$ vs. $B$ profile at around $B=0.27$T and $0.65$T. Similarly, in the case of magnetic modulation the dips
are around $B=0.22$T and $0.42$T. All these dips correspond to the Weiss oscillation minima (see Fig.~\ref{diff_con}).

\section{Combined effect of electric and magnetic modulation}\label{sec4}
\begin{figure}	
{\includegraphics[width=.47\textwidth,height=5.2cm]{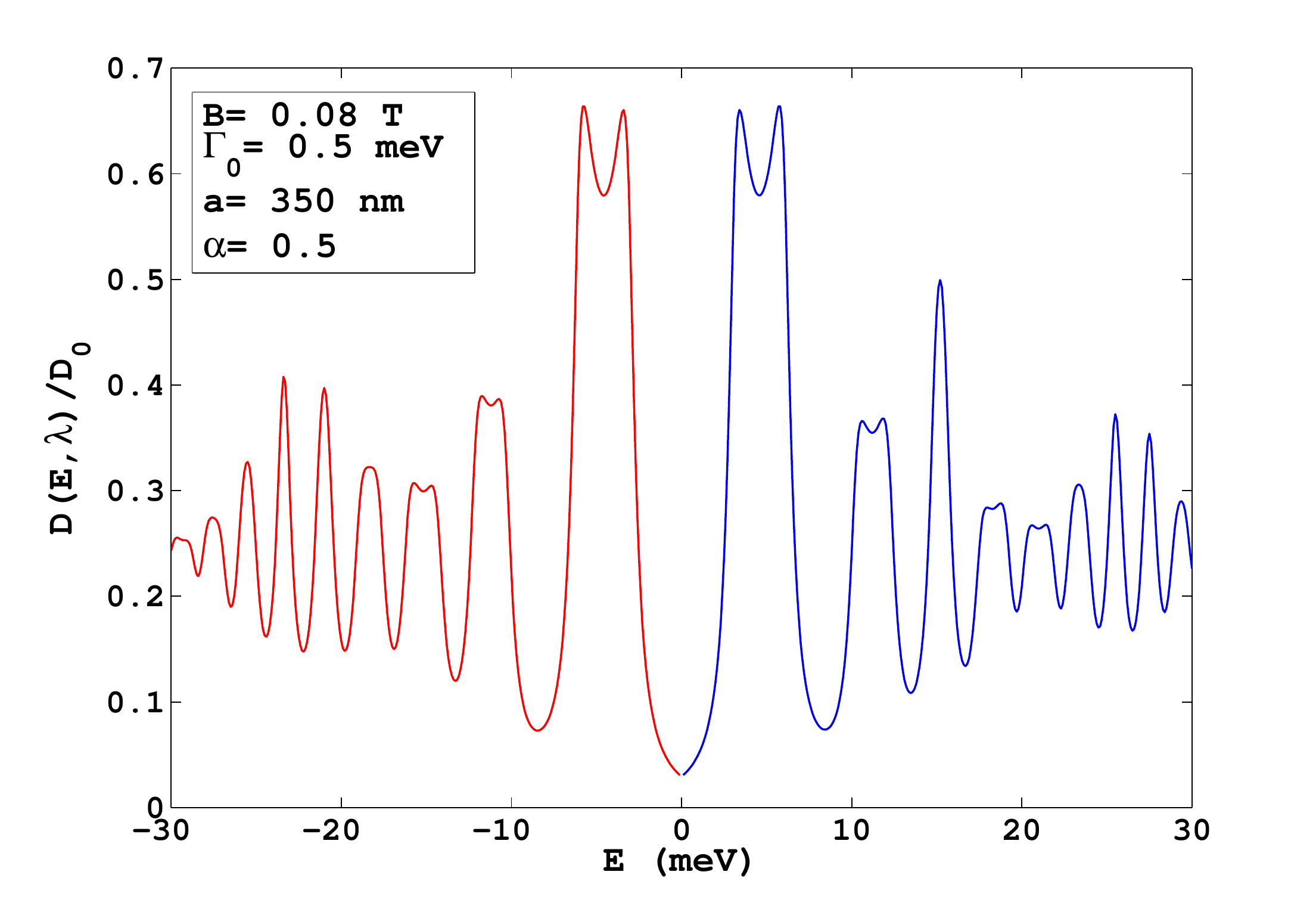}}
\caption{(Color online) Behavior of the modulated density of states in the presence of both types of modulations is shown for $K$
valley. The strengths of the electric and magnetic modulation are taken to be equal i.e., $V_e=V_m=1$ meV. The blue and red lines
represent the conduction and valence band, respectively.}
\label{dos}
\end{figure}
So far, we have considered the role of electric and magnetic modulations in the transport phenomena separately. In this 
subsection, we examine what happens when both electric and magnetic modulation are considered simultaneously. The presence of 
magnetic stripes or a superconductor on the top of the system generally induces an unwanted electric potential modulation too 
which motivated us to consider the effect of both modulations in the usual $2$DEG also~\cite{peeters}. In our present case, the 
presence of both types of modulations induces first order energy correction as,
\begin{equation}\label{breaking}
 \Delta E_{\xi,k_x}^{\lambda}=\frac{1}{2}[V_e F_{\xi}(u)+ \lambda\eta V_m R_{\xi}(u)]\cos(qy_0).
\end{equation}
Note that unlike the previous cases where either electric or magnetic modulation was considered, here the first order energy correction 
breaks the particle-hole symmetry around zero energy, as $\Delta E_{\xi,k_x}^+\neq\Delta E_{\xi,k_x}^-$. This is one of the main 
results of our study. Note that, this phenomenon could also be observed in Dirac material like graphene and silicene but it has not 
been pointed out in the literature to the best of our knowledge.
\begin{figure*}
{\includegraphics[width=.49\textwidth,height=6.6cm]{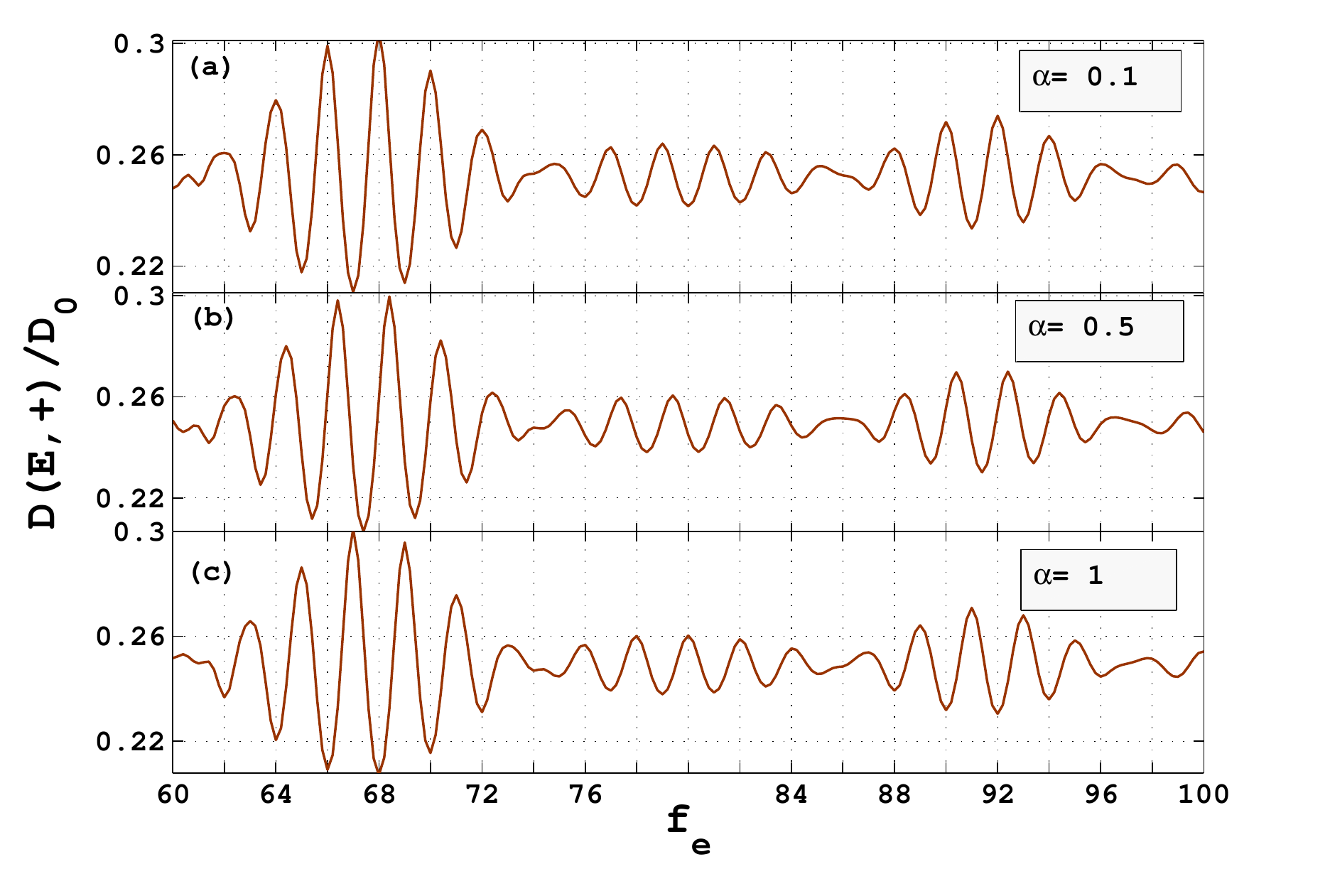}}
{\includegraphics[width=.49\textwidth,height=6.8cm]{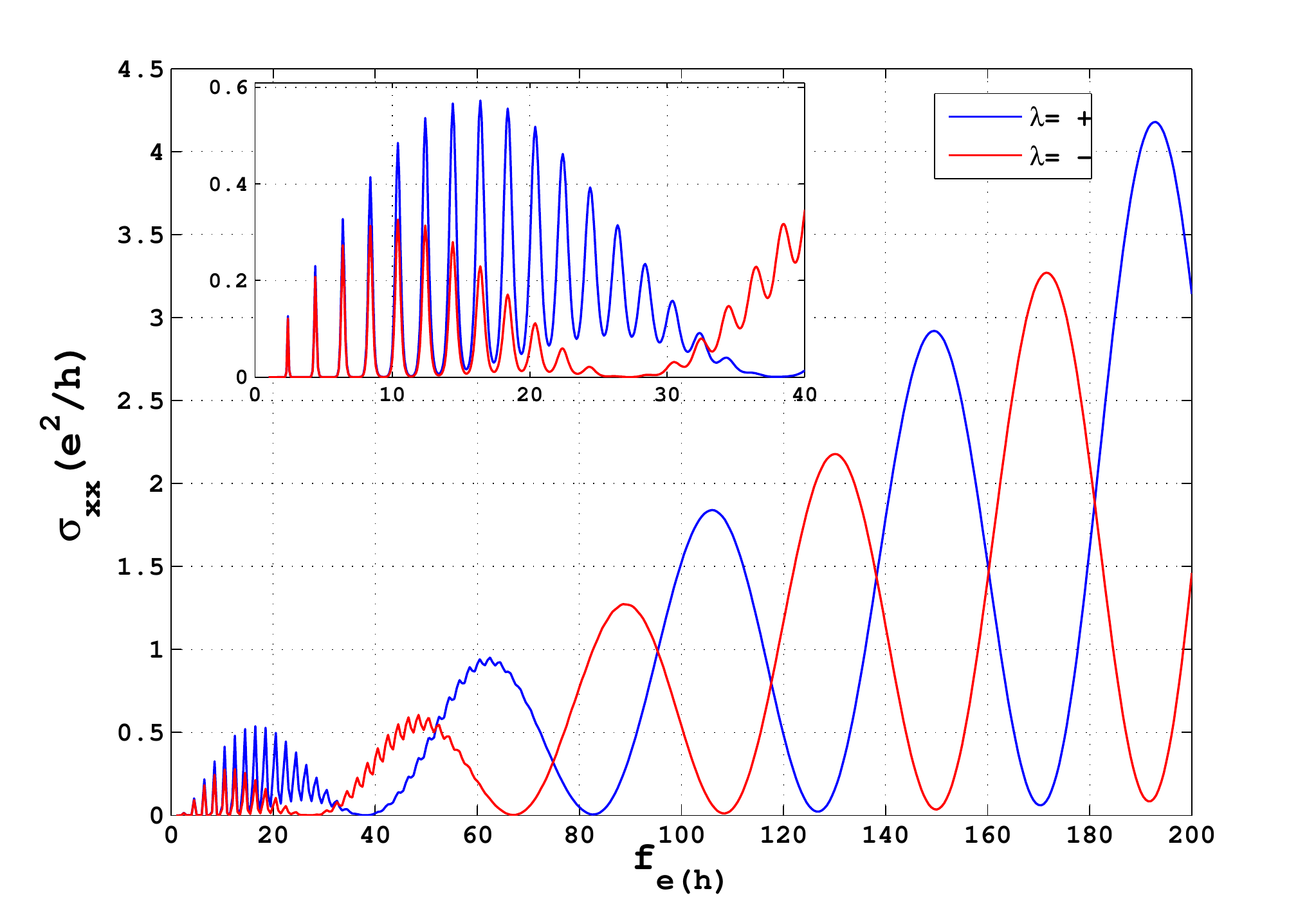}}
\caption{(Color online) Left panel: Modulated density of states in the $K$ valley vs. filling fraction $f_e$ ($=2 \pi n_e \hbar/eB$) 
for the conduction band is shown in the presence of both type of modulations for (a) $\alpha=0.1$, (b) $0.5$, and (c)$1$. The carrier density 
is taken here as $n_e=3 \times 10^{15}m^{-2}$ and the strength of modulation is taken to be $V_e=V_m=1.2$ meV. Right panel: Variation 
of diffusive conductivity with respect to the electron (hole) filling fraction $f_{e(h)}$ is plotted for the conduction (valence) band for the
intermediate value of $\alpha=0.5$ only.}
\label{dos_filling}
\end{figure*}
As stated before, the application of modulation manifests itself through modulated density of states (DOS) corresponding to the Landau 
level broadening. To gain more insight, in Fig.~\ref{dos} we plot the modulated DOS as a function of the energy $E$ in the
presence of the combined effects of electric and magnetic modulation. The DOS in the presence of modulation can be expressed in a Lorentzian 
distribution as
\begin{equation}
 \frac{D(E,\lambda)}{D_0}=\frac{(\hbar\omega_c)^2}{8\pi |E|}\sum_{n}\frac{1}{\pi}\int_0^{\pi}dc\frac{\Gamma_0}
 {(E-E_{\xi,c}^{\lambda})^2+\Gamma_0^2}
\end{equation}
with $D_0=2|E|/(\pi \hbar^2 v_F^2)$, $c=qy_0$ and 
$E_{\xi,c}^{\lambda}=\lambda \,\epsilon\sqrt{n+\chi_\eta}+V_0[F_{\xi}(u)+\lambda \eta R_{\xi}(u)]\cos(c)$ where $V_0=V_e=V_m$. Interestingly, 
the electron and hole band lost the symmetry after the inclusion of both modulations here. The heights of particular peaks in the electron 
(conduction) and hole (valence) band are not the same. This is clearer in the higher energy regime. We show the results for the $K$ valley. Similar 
particle-hole symmetry breaking in the DOS can be obtained in the $K^{\prime}$ valley too. 

Presence of both modulations together plays a vital role in transport phenomena of the usual $2$DEG as observed in experiments~\cite{filling_exp} 
followed by theoretical work by Shi \etal~\cite{shi}, especially in the behavior of the DOS and conductivity with the filling fraction. The peculiar 
phenomenon of odd-even filling fraction transition in the DOS has been explored. Motivated by this, we investigate the same in our $\alpha$-$\mathcal{T}_3$ 
model in order to reveal the role of Berry phase in this context. In the left panel in Fig.~\ref{dos_filling} we show the behavior of 
the modulated the DOS in the $K$ valley with respect to the electron filling fraction, $f_e$, for the conduction band only. Here, (a), (b) and (c) represent 
the results for $\alpha=0.1$, $0.5$ and $1$, respectively. We notice that there is a transition of the peak positions from even to odd filling 
fraction when we tune $\alpha$ from $0$ to $1$ as depicted in left panel of Fig.~\ref{dos_filling}. We show the results for a higher filling 
fraction which is inversely proportional to the magnetic field as $f_e=2 \pi n_e \hbar/eB$ for a fixed value of the carrier density. Now for 
zero or very small value of $\alpha$, the DOS shows some peaks at odd filling fraction \ie $f_e=77$, $79$, $81$, and $83$ and the rests are at even 
$f_e$. The DOS peaks at odd filling fraction indicates that the Landau levels are half-filled. On the contrary, when $\alpha$ is tuned to $1$ (dice 
lattice), we observe exactly the inverted picture \ie all peaks (dips) positions get inverted to dips (peaks) as displayed in (c) (Fig.~\ref{dos_filling} 
left panel). These two patterns correspond to the two limiting values of $\alpha$. There is a transition of the DOS peak positions from the even to odd 
or from odd to even filling fraction through beating nodes for a particular value of $\alpha$. In addition to this known feature, we notice a similar odd 
(even)-even (odd) transition through a smooth variation of $\alpha$ from $0$ to $1$. At intermediate $\alpha$, shown in (b) (Fig.~\ref{dos_filling}, left 
panel), each odd (even) peak gets shifted from its earlier positions for $\alpha=0$, and they are neither at even nor at odd filling fraction; rather they 
arise at fractional values of $f_e$. 

Now, we calculate the diffusive conductivity following an approach similar to that used in previous cases; it is given by,
\begin{equation}
 \sigma_{xx}^{\lambda,\eta}=\frac{e^2}{h}\frac{V_0^2\beta}{4\Gamma_0}u\sum_{n}f^{\lambda}_{\xi}(1-f^{\lambda}_{\xi})[F_{\xi}(u)+\lambda\eta R_{\xi}(u)]^2
\label{sigma_both}
 \end{equation}
where $\lambda$ is the band index. Both $\lambda$ and $\eta$ change their signs in the $K^{\prime}$ valley. Note that, unlike the previous 
cases of either electrically or magnetically modulated $\alpha$-$\mathcal{T}_3$ lattices, here the diffusive conductivity depends on the band index 
too [see Eq.(\ref{sigma_both})]. This is because of the band dependent group velocity which arises from the first order energy correction 
due to the modulation of both types. A similar band dependence of the group velocity also appears in the case of magnetic modulation 
[Eq.(\ref{v_mag})] but it does not affect the diffusive conductivity $\sigma_{xx}$, being proportional to $(v_x^m)^2$ as expressed in 
Eq.~(\ref{sigma_mag}). 

We plot the diffusive conductivity for the $K$ valley as a function of the filling fraction $f_{e(h)}$ for the conduction (valence) band, denoted by 
blue (red) curves in the right panel in Fig.~\ref{dos_filling}. The amplitudes of the Weiss oscillations for both band are enhanced due to 
the presence of both modulations. The diffusive conductivities for the conduction and valence bands are not similar to each other following the 
particle-hole asymmetry as shown in the DOS profile. There is a prominent phase difference between the Weiss oscillations for electron and hole 
bands. They change with the filling fraction as well as the magnetic field too. This feature of Weiss oscillation is in complete contrast to 
the usual $2$D gas because of the preservation of the particle-hole symmetry even in presence of both modulations. 
 
\begin{figure}
{\includegraphics[width=.47\textwidth,height=5.2cm]{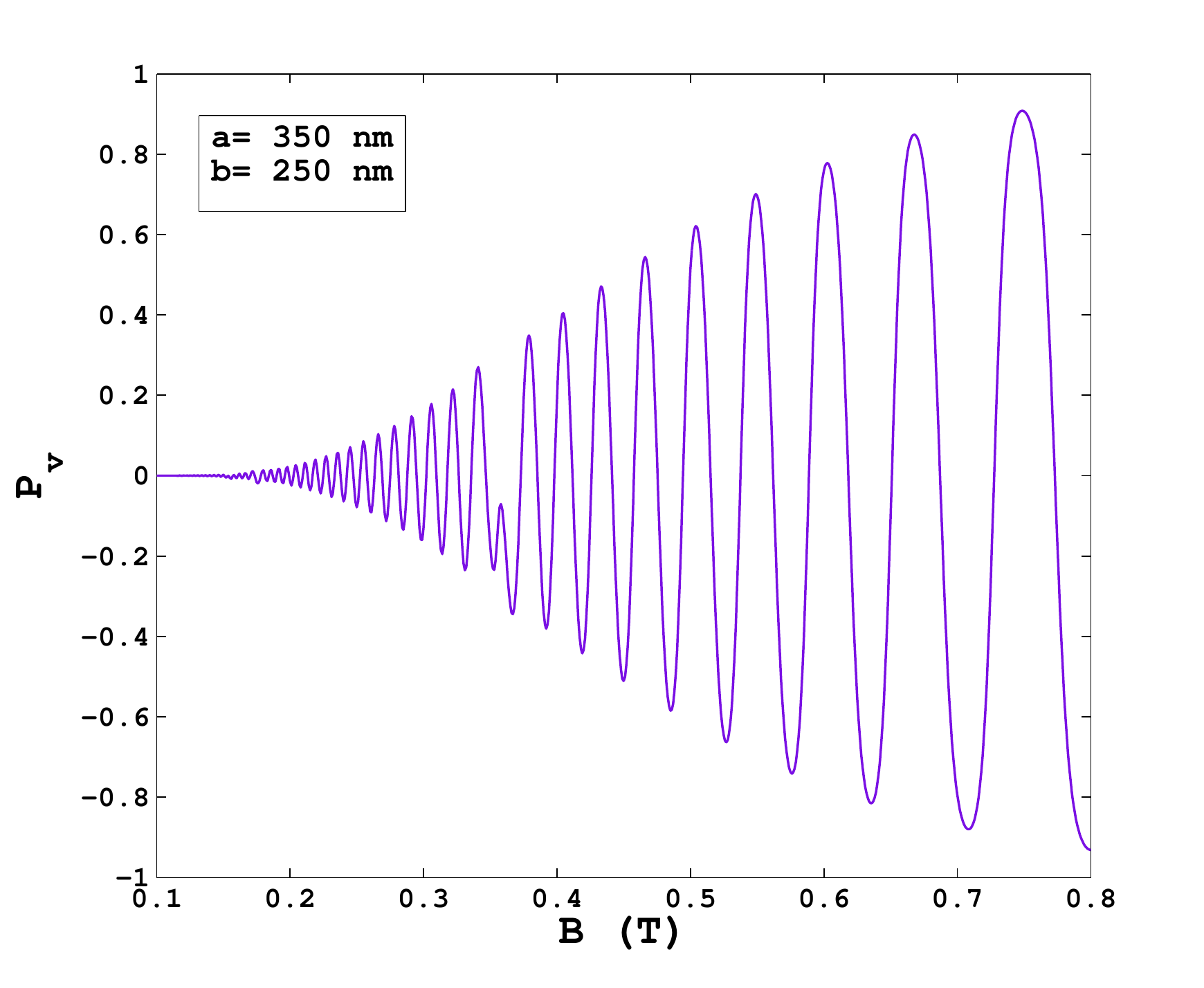}}
\caption{(Color online) Valley polarization in the presence of both modulations with $V_e=V_m=1$meV. The electric and magnetic modulation periods
are taken as $a$ and $b$, respectively.}
\label{pv_both}
\end{figure}
Finally, in Fig.~\ref{pv_both} we present the variation of valley polarization in the diffusive conductivity as a function of magnetic field
\begin{figure*}
\subfigure[]
{\includegraphics[width=.49\textwidth,height=5cm]{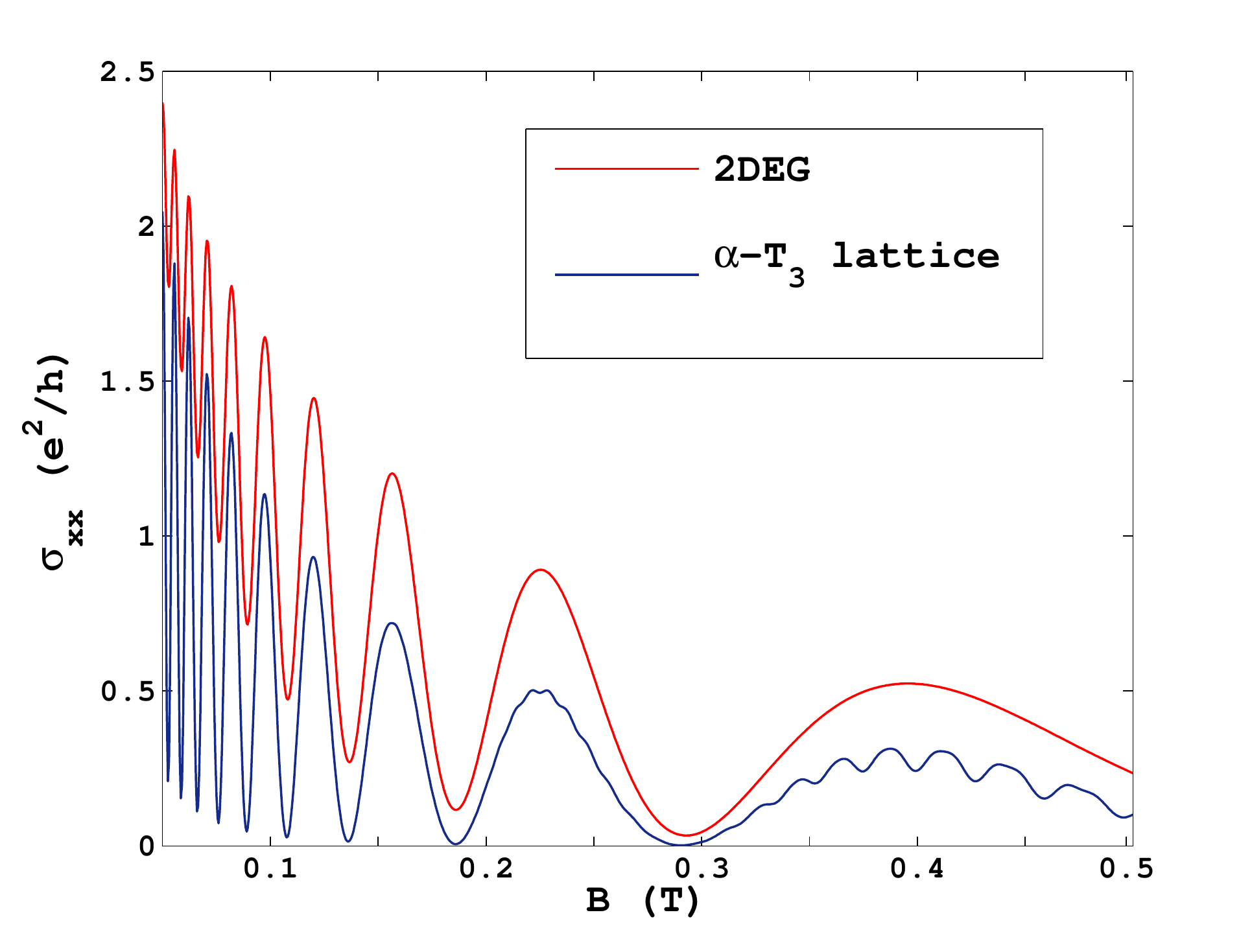}}
\subfigure[]
{\includegraphics[width=.49\textwidth,height=5cm]{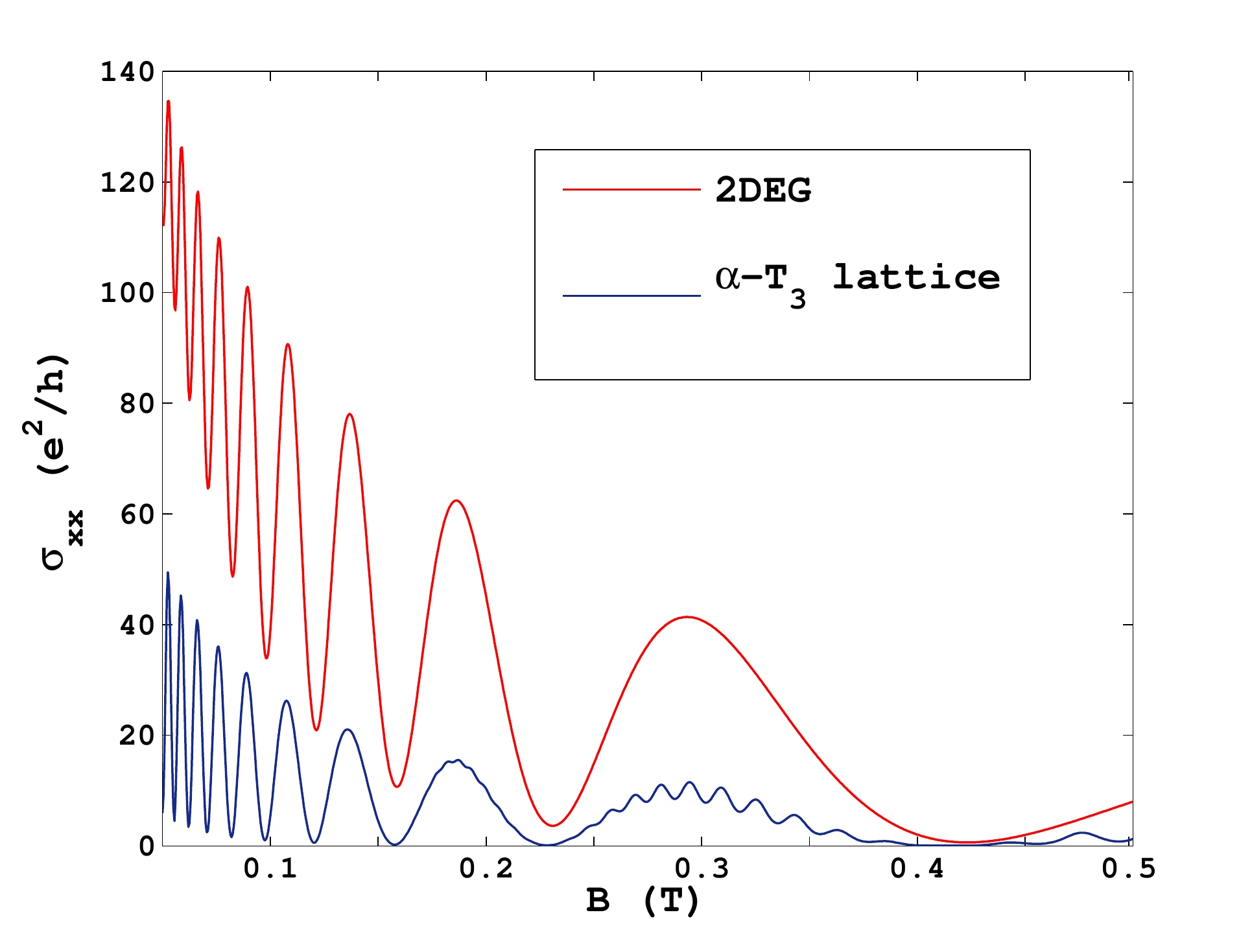}}
\caption{(Color online) Diffusive conductivity is plotted as a function of the magnetic field for the usual $2$DEG and $\alpha$-$\mathcal{T}_3$ 
lattice with $\alpha=0.5$ taking contributions from both the valleys: (a) electric modulation and (b) magnetic modulation. In the case of magnetic
modulation, the diffusive conductivity in the $\alpha$-$\mathcal{T}_3$ lattice is multiplied by a factor of $25$ in order to keep the scale the 
same as that for $2$DEG.}
\label{comprsn}
\end{figure*}
using Eq.(\ref{pola_defn}). In the presence of both modulations we observe that valley polarization naively remains unaltered from the case
where we consider either electric or magnetic modulation. Here, we set the $\alpha=0.5$ considering the strengths of both modulations to be 
equal to each other \ie $V_e=V_m=V_0$. However from the realistic point of view their strengths may not be equal to each other. Also the period 
of the modulations may be different. Note that, in Fig.~\ref{pv_both} we consider two different periods for the two different types of the 
modulation. The valley-polarization also does not change by appreciable amount with their relative periods.

\section{Comparison with other $2$D systems}

Now we draw a comparison between our present results and the existing results for other $2$D systems like the usual $2$DEG, graphene, and silicene. 

For the usual $2$DEG, the first-order energy correction has already been evaluated in Ref.~[\onlinecite{vasilo}] as,
\begin{equation}
\Delta E_{k_x}^{e}=V_{e}^{2d}e^{-\frac{u}{2}} 	L_{n}(u) \cos(q y_0)
\end{equation}
corresponding to the Landau levels $E_n=\hbar \omega_{2d} (n+1/2)$ with $\omega_{2d}=e B/m^*$, $m^*$ being the effective mass of the electrons
in $2$DEG. Using this correction the diffusive conductivity was evaluated in Ref.~[\onlinecite{vasilo}].
Note that, in the expression of the energy correction as well as in the diffusive conductivity for the $\alpha$-$\mathcal{T}_3$ lattice we have summation over three 
successive Laguerre polynomials whereas in $2$DEG only one Laguerre polynomial is involved in the expression (see Eq.~(\ref{Fexp}) and~(\ref{el_diff})). 
Moreover, in the $\alpha$-$\mathcal{T}_3$ system both the energy correction and diffusive conductivity are sensitive to the valley degree of freedom, whereas in a
conventional $2$DEG the issue of the valley degree of freedom is irrelevant.

Similarly, for magnetic modulation the first order energy correction reads~\cite{peeters},
\begin{equation}
\Delta E_{k_x}^{m}=V_{m}^{2d}e^{-\frac{u}{2}}\left[\left(\frac{1}{2}-\frac{n}{u}\right)L_n(u)+\frac{n}{u}L_{n-1}(u)\right]\sin(q y_0)
\end{equation}
and the corresponding diffusive conductivity can be found in Ref.~[\onlinecite{peeters}].
Compared with Eqs.~(\ref{Rexp}) and (\ref{sigma_mag}) it is clear that the summations over the Laguerre polynomial in both energy correction 
and diffusive conductivity corresponding to the two systems, $\alpha$-$\mathcal{T}_3$ and usual $2$DEG are different from each other.

In Fig.~\ref{comprsn} we plot the both the electric [Fig.~\ref{comprsn}(a)] and magnetic [Fig.~\ref{comprsn}(b)] modulation induced diffusive 
conductivity for our system as well as $2$DEG. For our numerical calculation the modulation strengths, electric as well as magnetic, in both 
the systems are fixed to $1$ meV. We find that the amplitude of Weiss oscillations in electrically modulated $2$DEG is higher than that in 
the $\alpha$-$\mathcal{T}_3$ model as depicted in Fig.~\ref{comprsn}(a). Also, they are naively in the same phase as each other. In the case of our system 
SdH oscillation starts to superimpose over Weiss oscillation under a relatively low magnetic field (B$\sim0.3$T) regime compared to that in 
$2$DEG (B$>0.5$T). The reason is ascribed to the behavior of Landau levels with the magnetic field. In Dirac material the Landau level 
$E\propto \sqrt{B}$ whereas it is proportional to $B$ in the usual $2$DEG. Additionally, the presence of the Berry phase in the $\alpha$-$\mathcal{T}_3$ 
model makes the diffusive conductivity behave differently in the two valleys leading towards the valley polarization. In the case of magnetic 
modulation, as displayed in Fig.~\ref{comprsn}(b), the order of magnitude of the diffusive conductivity is strongly suppressed in $\alpha$-$\mathcal{T}_3$ 
in comparison to that in the usual $2$DEG. This type of damped oscillation observed in the $\alpha$-$\mathcal{T}_3$ lattice with respect to $2$DEG is similar as 
explored in graphene~\cite{tahir_gra}. However, conductivity oscillation in usual $2$DEG and $\alpha$-$\mathcal{T}_3$ are almost in phase. The most 
remarkable differences between these two systems are appearance of valley polarization and particle-hole symmetry breaking in $\alpha$-$\mathcal{T}_3$ 
lattice in comparison to the usual $2$DEG. The valley polarization in $\alpha$-$\mathcal{T}_3$ lattice is attributed to the valley dependent mass
term ($\chi_\eta$) in Landau levels. The total energy correction due to the application of both modulation does not depend
on the band index in usual $2$DEG, whereas it is strongly dependent on the band index ($\lambda$) in each valley of the
$\alpha$-$\mathcal{T}_3$ lattice [see Eq.~\ref{breaking}] leading towards the breaking of particle-hole symmetry.

Now we look into how the $\alpha$-$\mathcal{T}_3$ system differs  from graphene in diffusive conductivity. In our system, the first order energy
correction [see Eq.(\ref{corr_elec})] in the electric modulation case is weakly sensitive to the valley index and contains the summation 
of three successive Laguerre polynomials. On the other hand, there are summation over two successive Laguerre polynomials in case of 
graphene without any coefficient (see Ref~\onlinecite{matulis_gra}). A similar difference in the form of the energy correction in both systems
can also be observed in the magnetic modulation case. The results for $\alpha=0$ corresponds to graphene, as shown in Fig.~\ref{diff_con},
having features similar to those with the $\alpha$-$\mathcal{T}_3$ lattice. However, the $\alpha$-$\mathcal{T}_3$ lattice exhibits the valley polarization in diffusive
conductivity with an increase in the magnetic field, which is absent in graphene due to the valley degeneracy in Landau levels. Similarly
to the $\alpha$-$\mathcal{T}_3$ lattice, the phenomenon of particle-hole symmetry breaking can also be obtained in graphene but it has not yet been 
pointed out in the literature to the best of our knowledge. 

A similar valley polarization has also been observed in electrically modulated silicene~\cite{vasi_sili}, even under a low magnetic field regime, 
but in the presence of a staggered potential between two sublattice planes. Weiss oscillation in magnetically modulated silicene has not been 
studied so far, to the best of our knowledge. 

\section{Summary and Conclusion}\label{sec5}

To summarize, in this article we have theoretically studied magneto-transport properties of a spatially modulated $\rm \alpha$-$\mathcal{T}_3$ lattice. 
Both electric and magnetic modulation, individually as well as simultaneously, have been considered here. Using the Kubo formula, based 
on linear response theory~\cite{kubo2}, we have obtained a modulation-induced additional contribution to the longitudinal conductivity, \ie 
diffusive conductivity. The unique feature of the $\rm \alpha$-$\mathcal{T}_3$ lattice is the tunable Berry phase, which ranges from $0$ to $\pi$. We 
have exploited this feature in order to reveal how it affects the quantum transport properties of the modulated $\rm \alpha$-$\mathcal{T}_3$ lattice.
The presence of modulation imparts a non-zero drift velocity to the electrons which is oscillatory with magnetic field, and leads to the rise 
of Weiss oscillation in the electrical conductivity signal with the magnetic field. We have noticed that a sizable valley polarization appears 
in the diffusive conductivity depending on the magnetic field. With the increase in the magnetic field, SdH oscillations start to superimpose 
over Weiss oscillations and valley polarization becomes much stronger and oscillates rapidly due to the strong dependence of the Landau levels 
on the Berry phase. To understand valley polarization with an increase in the magnetic field, we have derived an analytical form of the density 
of states and used it to get an approximate analytical expression for the diffusive conductivity. The signature of the Berry phase in diffusive 
conductivity exclusively enters through the modification of Landau levels and the corresponding states of the system. On the contrary, we have 
also checked that in the absence of a magnetic field and modulation, the diffusive conductivity is independent of the Berry phase. 

We have compared our results in for the $\alpha$-$\mathcal{T}_3$ lattice with the existing results in the literature for the usual $2$DEG and other Dirac 
material such as graphene. We have observed that Weiss oscillation in the $\alpha$-$\mathcal{T}_3$ lattice is almost in the same phase as the usual 2DEG 
except for an amplitude mismatch. The origin of the amplitude mismatch is due to the Dirac nature of band dispersion in the $\alpha$-$\mathcal{T}_3$ lattice 
or graphene, which was already pointed out in Ref.[\onlinecite{matulis_gra}]. However, the most exciting physics in the $\alpha$-$\mathcal{T}_3$ lattice is 
the appearance of valley polarization and particle-hole symmetry breaking in comparison to usual 2DEG. On the other hand, though graphene and the 
$\alpha$-$\mathcal{T}_3$ lattice both exhibit a Dirac-like band dispersion, the presence of additional atoms at the center of each hexagon in the $\alpha$-$\mathcal{T}_3$ 
lattice causes a valley polarization in diffusive conductivity which is absent in graphene. Moreover, graphene and the $\alpha$-$\mathcal{T}_3$ lattice both 
exhibit particle-hole symmetry breaking under the influence of both modulations. Very recently, the Carbotte group has shown that a small 
asymmetry in the energy band  can be very sensitive to magneto-optical excitation~\cite{carbotte1,carbotte2}. Therefore, we conclude that the combination 
of both modulations can be used as a tool to break the particle-hole symmetry for manipulation of the valley degree of freedom in optical devices. Moreover, 
we have explored a modulation induced transition of odd (even)-to-even (odd) filling fraction corresponding to DOS peaks with the variation of 
$\alpha$.

As far as the practical realization of this $\rm \alpha$-$\mathcal{T}_3$ lattice is concerned, it can be naturally formed by growing a trilayer structure of 
cubic lattices in the (111) direction as shown by Wang \etal~\cite{fwang}. On the other hand, Bercioux \etal~have proposed an experimental 
set up to realize this lattice by confining ultra-cold atoms to an optical lattice~\cite{urban}. On the other hand, periodic modulation can be 
engineered in several ways. For example, Winkler \etal~\cite{kotha} have used an array of biased metallic strips on the surface of a $2$D 
electronic system to achieve electric modulation. Magnetic modulation can be achieved by placing a few patterned ferromagnets or a superconductor 
on the surface of the $2$D material~\cite{mag_exp1,mag_exp2,mag_exp3}.

\begin{acknowledgements}
We specially acknowledge Arijit Saha for stimulating discussions and careful reading of the manuscript. SFI also acknowledges Tarun 
Kanti Ghosh for useful discussions. PD thanks Science and Engineering Research Board (SERB), Department of Science 
and Technology (DST), India for the financial support through National Post-Doctoral Fellowship (File No. PDF/2016/001178). We 
acknowledge Arun M. Jayannavar for his encouragement and support.
\end{acknowledgements}

\end{document}